\documentclass[twocolumn,preprintnumbers]{revtex4}

\usepackage{graphicx}
\usepackage{epsf}
\usepackage{amsmath,amssymb}
\usepackage{color}
\newcommand{\bvec}{\boldsymbol}

\newcommand{\Mg}{{}^{24}\textrm{Mg}}

\usepackage{ulem}
\begin{document}
\preprint{KUNS-2839, NITEP 79}
\title{Probing negative-parity states of $^{24}$Mg 
probed via proton and alpha inelastic scattering}

\author{Yoshiko Kanada-En'yo}
\affiliation{Department of Physics, Kyoto University, Kyoto 606-8502, Japan}
\author{Kazuyuki Ogata} %\email{kazuyuki@rcnp.osaka-u.ac.jp}
\affiliation{Research Center for Nuclear Physics (RCNP), Osaka University,
  Ibaraki 567-0047, Japan}
\affiliation{Department of Physics, Osaka City University, Osaka 558-8585,
  Japan}
\affiliation{
Nambu Yoichiro Institute of Theoretical and Experimental Physics (NITEP),
   Osaka City University, Osaka 558-8585, Japan}

\begin{abstract}
\begin{description}
\item[Background:]
The band structure of the negative-parity states of $^{24}$Mg has not yet been clarified. 
The $K^\pi=0^-$, $K^\pi=1^-$, and $K^\pi=3^-$ bands have been suggested, but 
the assignments have been inconsistent between experiments and theories. 
\item[Purpose:]
Negative-parity states of $^{24}$Mg are investigated by microscopic structure and reaction 
calculations via proton and alpha inelastic scattering to clarify the band assignment for the observed negative-parity spectra.
\item[Method:]
The structure of $^{24}$Mg  was calculated using the antisymmetrized molecular dynamics~(AMD).  
Proton and alpha inelastic reactions were calculated using microscopic coupled-channel (MCC) calculations 
by folding the Melbourne $g$-matrix $NN$ interaction 
with the AMD densities of $^{24}$Mg.
\item[Results:]
The member states of the $K^\pi=0^+$, $K^\pi=2^+$, $K^\pi=0^-$, $K^\pi=1^-$, and $K^\pi=3^-$ bands of $^{24}$Mg 
were obtained through the AMD result. In the MCC+AMD results for proton and alpha elastic and inelastic cross sections, reasonable agreements were obtained with existing data, except in the case of the $4^+_1$ state. 
\item[Conclusions:]
The $3^-$ state of the $K^\pi=3^-$ band and the $1^-$ and $3^-$ states of the $K^\pi=0^-$ 
bands were assigned to the $3^-_1$(7.62~MeV), $1^-_1$(7.56~MeV), and $3^-_2$(8.36~MeV) states, respectively. 
The present AMD calculation is the first microscopic structure calculation to
reproduce the energy ordering of the $K^\pi=0^-$, $K^\pi=1^-$, and $K^\pi=3^-$ bands of $^{24}$Mg.
\end{description}
\end{abstract}

\maketitle

\section{Introduction}
The band structure of $^{24}$Mg has been studied via electromagnetic 
transitions~\cite{Branford:1975ciy,Fifield:1979gfv,Keinonen:1989ltz}, and 
intensively investigated through inelastic scattering 
of various probes, including electrons~\cite{Horikawa:1971oau,Horikawa1972,Nakada:1972,Johnston_1974,Li:1974vj,Zarek:1978cvz,Zarek:1984fm}, 
pions~\cite{Blanpied:1990vd}, 
nucleons~\cite{Haouat:1984zz,Rush:1967zwr,Rush:1968qwc,Zwieglinski:1978zza,Zwieglinski:1978zz,Lombard:1978zz,Kato:1985zz,Horowitz:1969eso,Ray:1979zza,Blanpied:1979im,DeLeo:1981zz,Amos:1984aph}, $^3$He~\cite{Griffiths:1967hrr,VanDerBorg:1979pzv},
 and $\alpha$~\cite{Naqib1968,Rebel:1972nip,VanDerBorg:1979pzv,VanDerBorg:1981qiu,Adachi:2018pql}.

For inelastic hadron scattering off $\Mg$, 
detailed reaction analyses have been performed using
the distorted-wave born approximation (DWBA) and coupled-channel (CC) calculations. 
Reaction calculations using the phenomenological potentials of collective models 
have succeeded in describing the cross sections 
of low-lying positive-parity states in the $K^\pi=0^+$ ground- and $K^\pi=2^+$ side-bands
(other than the $4^+_1$ state), and have also suggested deformations, including triaxial and hexadecapole 
shapes for $^{24}$Mg. 

For low-lying negative-parity states of $\Mg$, member states in the $K^\pi=3^-$ and $K^\pi=0^-$ bands and 
candidates for the $K^\pi=1^-$ band have been reported by measurements of
the $\gamma$-decays~\cite{Fifield:1979gfv}, 
but the description of inelastic cross sections via reaction calculations has proven unsatisfactory, 
and the band assignments of negative-parity spectra have not yet been confirmed. 
In experiments with electron inelastic scattering, 
various behaviors of the form factors have been observed 
for two $3^-$ states: the $3^-_1$(7.62~MeV) state of the $K^\pi=3^-$ band and the $3^-_2$(8.36~MeV) state of 
the $K^\pi=0^-$ band~\cite{Zarek:1984fm}. 
A structure calculation using the open-shell random-phase approximation (RPA) 
has predicted two types of particle-hole excitations for the two $3^-$ states and 
qualitatively described only the first peak of the observed form factors,
but not the second peak of the $3^-_2$(8.36 MeV) state.  
For the band-head  $1^-_1$(7.56 MeV) state of the $K^\pi=0^-$ band and the $1^-_2$(8.44 MeV) state
of the $K^\pi=1^-$ bands, 
no calculation has yet succeeded in describing either form factors or inelastic hadron scattering.
Moreover, high-quality electron and hadron inelastic scattering data
for the $1^-$ states have been limited in quantity because it is generally difficult to 
resolve the $1^-_1$(7.56 MeV) and  $1^-_2$(8.44 MeV) states from the 
highly populated $3^-_1$(7.62 MeV) and $3^-_2$(8.36 MeV) spectra existing closely to 
the weak $1^-$ spectra in inelastic scattering.

Recently, the structure of the negative-parity states of $\Mg$ has been investigated by 
microscopic calculations using quasiparticle RPA~\cite{Nesterenko:2017rcc,Nesterenko:2019dnt} 
and antisymmetrized molecular dynamics~(AMD)~\cite{Kimura2012,Chiba:2019dap}. 
These structural studies have predicted low-lying isoscalar dipole excitations 
in the $K^\pi=0^-$ and $K^\pi=1^-$ bands 
and discussed the importance of the triaxial deformation and cluster structures of $\Mg$ 
for negative-parity excitations.
However, the predicted energy spectra of the negative-parity bands 
have been inconsistent with the experimental band assignment presented in Ref.~\cite{Fifield:1979gfv}, 
and the negative-parity band structure of $\Mg$ remains an open problem.

In the present paper, we aim to investigate the structure of low-lying states of $\Mg$ 
via analyses of inelastic electron, proton, and $\alpha$ scattering. 
Our main interest is in the low-lying 
$1^-$ and $3^-$ states of the negative-parity bands. 
In general, inelastic electron and high-energy-proton scattering directly detects 
the transition densities of excitations from the ground state, 
whereas the $\alpha$ scattering can sensitively probe the transitions 
at the outer surface region of target nuclei 
rather than the interior region, because of the strong absorbing $\alpha$-nucleus potentials. 
Moreover, low-energy proton and $\alpha$ scattering 
may contain information about in-band transitions via CC effects and can, in principle, 
be used as probes for the band assignment.
As for the structural inputs of $\Mg$, 
the existing $(e,e')$ data have shown a strong state dependence of the form factors 
indicating that simple collective models do not work in describing the transition densities 
of inelastic transitions. 
Furthermore, exotic deformations beyond axial symmetric-quadrupole deformation, 
cluster structures, and configuration mixing are expected to be important in 
the low-lying states, including the ground state of the $\Mg$ system. 

To achieve this aim, we apply 
the AMD method~\cite{KanadaEnyo:1994kw,KanadaEnyo:1995tb,KanadaEn'yo:1998rf,KanadaEn'yo:2012bj} 
for the structure calculation of $\Mg$ 
and perform microscopic coupled-channel (MCC) calculations 
of proton and $\alpha$ scattering. 
In the MCC calculations, the diagonal and transition densities of the target nuclei 
obtained with microscopic structure models are utilized as inputs of the 
CC reaction calculations in microscopic folding models, wherein the
nucleon-nucleus and $\alpha$-nucleus potentials are constructed by folding the effective $NN$ interactions. 
In our previous 
studies~\cite{Kanada-Enyo:2019prr,Kanada-Enyo:2019qbp,Kanada-Enyo:2019uvg,Kanada-Enyo:2020zpl,Kanada-Enyo:2020goh,Ogata:2020umn}, 
we have applied the MCC calculations to proton and $\alpha$ scattering off
various target nuclei in the $p$- and $sd$-shell regions
using the AMD densities and  the Melbourne $g$-matrix $NN$ interaction \cite{Amos:2000}. 
%The Melbourne $g$-matrix $NN$ interaction, which is an effective $NN$ interaction in a nuclear medium
%based on a bare $NN$ interaction of the Bonn B potential~\cite{Mac87}, 
%contains energy and density dependencies and does not require phenomenological adjustment of the interaction %parameters, owing to its fundamental derivation. 
We have presented successful results of the MCC+AMD approach 
for the $(p,p')$ and $(\alpha,\alpha')$ cross sections of various excited states.

In this paper, 
we first calculate the structure of $\Mg$ with 
variation after parity and total-angular-momentum projections (VAP) 
in the AMD framework. 
The electromagnetic data (including transition strengths and electron scattering) are used to 
test the AMD result for the structural inputs. In particular, we compare the 
calculated transition strengths and form factors with the experimental data 
to check the assignment of predicted states to the experimental energy levels.
To use the reaction calculations, 
we renormalize the AMD transition density to fit the electric-transition strengths, 
so as to reduce the model ambiguity of the structural inputs. 
We then apply the MCC approach to proton and $\alpha$ scattering 
off $\Mg$ with the Melbourne $g$-matrix 
$NN$ interaction using the AMD densities of $\Mg$.
By analyzing these structure and reaction calculations, 
we can investigate the structure and transition properties of the ground~($K^\pi=0^+$), $K^\pi=2^+$, 
$K^\pi=3^-$, $K^\pi=0^-$, and $K^\pi=1^-$ bands.

The rest of this study is organized as follows. In Sec.~\ref{sec:method},
the frameworks for the AMD calculation for $\Mg$ and 
for the MCC approach to proton and $\alpha$ scattering off $\Mg$
are explained. 
The AMD results for the structural properties are described in Sec.~\ref{sec:results1}, while
Sec.~\ref{sec:results2} presents the proton- and $\alpha$-scattering results.  
Finally, a summary is given in Sec.~\ref{sec:summary}.

\section{Method} \label{sec:method} 
For the structure calculation of $\Mg$, 
we apply a VAP version of AMD, which 
is sometimes called AMD+VAP (though we use the name AMD in the present paper).
This method has been applied for structural studies of various nuclei including $^{12}$C and 
neutron-rich Be isotopes~\cite{KanadaEn'yo:1998rf,Kanada-Enyo:1999bsw,Kanada-Enyo:2003fhn},
and has also been used in the MCC+AMD calculation for reaction studies of proton and $\alpha$ scattering 
in Refs.~\cite{Kanada-Enyo:2019prr,Kanada-Enyo:2019qbp,Kanada-Enyo:2019uvg,Kanada-Enyo:2020zpl,Kanada-Enyo:2020goh}.
The calculational procedures of the present structure and reaction calculations 
are almost the same as the MCC+AMD calculation for $^{20}$Ne 
in Ref.~\cite{Kanada-Enyo:2020goh}. 
For details, the reader is referred to those papers and the references contained therein.

\subsection{AMD calculations for $\Mg$}

In the AMD framework, an $A$-nucleon wave function 
is expressed by the Slater determinant of 
single-nucleon Gaussian wave functions as
\begin{eqnarray}
 \Phi_{\rm AMD}({\bvec{Z}}) &=& \frac{1}{\sqrt{A!}} {\cal{A}} \{
  \varphi_1,\varphi_2,...,\varphi_A \},\label{eq:slater}\\
 \varphi_i&=& \phi_{{\bvec{X}}_i}\chi_i\tau_i,\\
 \phi_{{\bvec{X}}_i}({\bvec{r}}_j) & = &  \left(\frac{2\nu}{\pi}\right)^{3/4}
\exp\bigl[-\nu({\bvec{r}}_j-\bvec{X}_i)^2\bigr],
\label{eq:spatial}\\
 \chi_i &=& (\frac{1}{2}+\xi_i)\chi_{\uparrow}
 + (\frac{1}{2}-\xi_i)\chi_{\downarrow}.
\end{eqnarray}
Here, ${\cal{A}}$ is the antisymmetrizer, and  $\varphi_i$ is
the $i$th single-particle wave function, written as a product of 
the spatial ($\phi_{{\bvec{X}}_i}$), spin ($\chi_i$), and isospin ($\tau_i$)
wave functions, where $\tau_i$ is fixed to be a proton or a neutron.
The $\nu$ value of the width parameter is common for all single-nucleon Gaussians
and is chosen to be $\nu=0.16$ fm$^{-2}$, which reproduces the root-mean-square radius of $^{16}$O in the 
harmonic oscillator $p$-shell closed configuration. 
Parameters 
${\bvec{Z}}\equiv
\{{\bvec{X}}_1,\ldots, {\bvec{X}}_A,\xi_1,\ldots,\xi_A \}$ (representing  
the Gaussian centroid positions and nucleon-spin orientations of the single-particle wave functions) 
are treated as variational 
parameters and determined by the energy optimization for each $J^\pi$ state of $\Mg$.
Energy variation is performed after the parity and total-angular-momentum projections
so as to minimize the energy expectation value 
$E=\langle \Psi|{\hat H}|\Psi\rangle /\langle \Psi|\Psi\rangle$ for
$\Psi=P^{J\pi}_{MM'}\Phi_{\rm AMD}({\bvec{Z}})$ as projected from the AMD wave function 
with the parity and total-angular-momentum projection operator $P^{J\pi}_{MM'}$. 

The VAP is performed for $(J^\pi, M')=(0^+,0),(2^+,0),(3^+,2)$, and $(4^+,0)$ 
to obtain the member states of the $K^\pi=0^+$ and $K^\pi=2^+$ bands. 
For negative-parity states, VAPs are performed using 
$(J^\pi, M')=(1^-,0),(2^-,1),(3^-,0),(3^-,3),(4^-,3)$ and $(5^-,0)$, 
and the member states of the $K^\pi=3^-$,  $K^\pi=0^-$, and $K^\pi=1^-$ bands
are obtained. 
Here $M'$ is the quanta of the 
$Z$ component $J_Z$ of the total-angular-momentum in the body-fixed frame but does not 
necessarily indicate the $K$ quanta defined for the principal axis of the intrinsic state, 
because there is no constraint upon the orientation 
of the intrinsic deformation in the energy variation. This means that the principal axis of the intrinsic 
deformation can, in principle, be tilted from the $Z$ axis. Indeed, VAP calculations 
with $(J^\pi, M')=(3^-,0), (3^-,3), (4^-,3)$ and $(5^-,0)$ yield the dominant configurations 
for the $K^\pi=3^-$ band, whereas, in VAP calculations with 
$(J^\pi, M')=(1^-,0)$, two kinds of configurations corresponding to the $K^\pi=0^-$ and $K^\pi=1^-$ 
bands are obtained as local minima. 

After the VAP calculations, we obtain the optimized sets, $\bvec{Z}^{(m)}$, for
the intrinsic configurations $\Phi_{\rm AMD}(\bvec{Z}^{(m)})$, which are labeled by $m$ for each parity
as $m=1,\ldots,4$ for the positive-parity states and $m=1,\ldots,7$ for the negative-parity states.   
To obtain the final wave functions for the $J^\pi$ states of $\Mg$, 
the obtained configurations are superposed by diagonalizing the Hamiltonian and norm matrices using the 
basis wave functions $P^{J\pi}_{MM'}\Phi_{\rm AMD}({\bvec{Z}}^{(m)})$, as 
projected from the obtained configurations.
Such diagonalization is performed for $M'$ and $m$, which correspond to $K$-mixing 
and  configuration $(m)$ mixing, respectively. 

The effective nuclear interactions used in the present AMD calculation
are the same as those in Refs.~\cite{KanadaEn'yo:1998rf,Kanada-Enyo:2019prr,Kanada-Enyo:2019qbp,Kanada-Enyo:2019uvg,Kanada-Enyo:2020zpl};
they are the  MV1 (case 1) central force \cite{TOHSAKI} with the parameters $(b,h,m)=(0,0,0.62)$
and the spin-orbit term of the G3RS force \cite{LS1,LS2} with strength parameters
$u_{I}=-u_{II}=3000$ MeV.
The Coulomb force is also included.

\subsection{MCC calculation of proton and $\alpha$ scattering off  $\Mg$}

The elastic and inelastic cross sections of  proton and $\alpha$ scattering off $\Mg$ are calculated via
MCC+AMD. The nucleon-nucleus potentials are constructed 
in a microscopic folding model, in which the diagonal and coupling potentials are calculated
by folding the Melbourne $g$-matrix $NN$ interaction \cite{Amos:2000} with the AMD 
diagonal and transition densities of $\Mg$.
The $\alpha$-nucleus potentials are obtained by folding the calculated
nucleon-nucleus potentials with 
an $\alpha$ density in an extended nucleon-nucleus
folding (NAF) model~\cite{Egashira:2014zda}.

The Melbourne $g$ matrix is an effective $NN$ interaction derived from a bare $NN$ interaction of 
the Bonn B potential~\cite{Mac87}. It contains energy and density dependencies 
with no adjustable parameter, and can be well applied to a systematic description of 
elastic and inelastic proton scattering off various nuclei
at energies of $E_p=$40--300~MeV ~\cite{Amos:2000,Minomo:2009ds,Toyokawa:2013uua,Minomo:2017hjl,Kanada-Enyo:2019uvg,Kanada-Enyo:2020zpl,Kanada-Enyo:2020goh}
and also of elastic and inelastic $\alpha$ scattering at energies of $E_\alpha=$100--400~MeV
\cite{Egashira:2014zda,Minomo:2016hgc,Kanada-Enyo:2019prr,Kanada-Enyo:2019qbp,Kanada-Enyo:2020zpl,Kanada-Enyo:2020goh}.
In the present reaction calculation, 
the spin-orbit term of the proton-nucleus potential is not account for to avoid complexity, 
as in Refs.~\cite{Kanada-Enyo:2019uvg,Kanada-Enyo:2020zpl}. 

As  structure inputs for the target nucleus, the diagonal $(\rho(r))$ and transition 
$(\rho^\textrm{tr}(r))$ densities of $\Mg$, as 
obtained by the AMD calculation, are used. 
To reduce the model ambiguity from the structure calculation, 
the theoretical-transition densities obtained by the AMD calculation
are renormalized by the factor $f^\textrm{tr}$
as $\rho^\textrm{tr}(r)\to f^\textrm{tr}\rho^\textrm{tr}(r)$
to fit the electromagnetic transition strengths
or $(e,e')$ data. If there are no data 
concerning the transition strength,  
the original AMD transition densities are used without renormalization. 

The $J^\pi=0^+_1$, $1^-_{1,2}$, $2^+_{1,2}$, $3^-_{1,2,3}$, and $4^+_{1,2}$ 
states of $\Mg$, and 
all $\lambda\le 4$ transitions between them are included 
in the CC calculation.
For the excitation energies of $\Mg$, experimental values are adopted. 

\section{Structure of $^{24}$Mg} \label{sec:results1}
\subsection{Band structure of $^{24}$Mg}

%%%%%%%%%%%%%%%%%%%%%%%%%%%%%%
\begin{figure}[!h]
\includegraphics[width=8 cm]{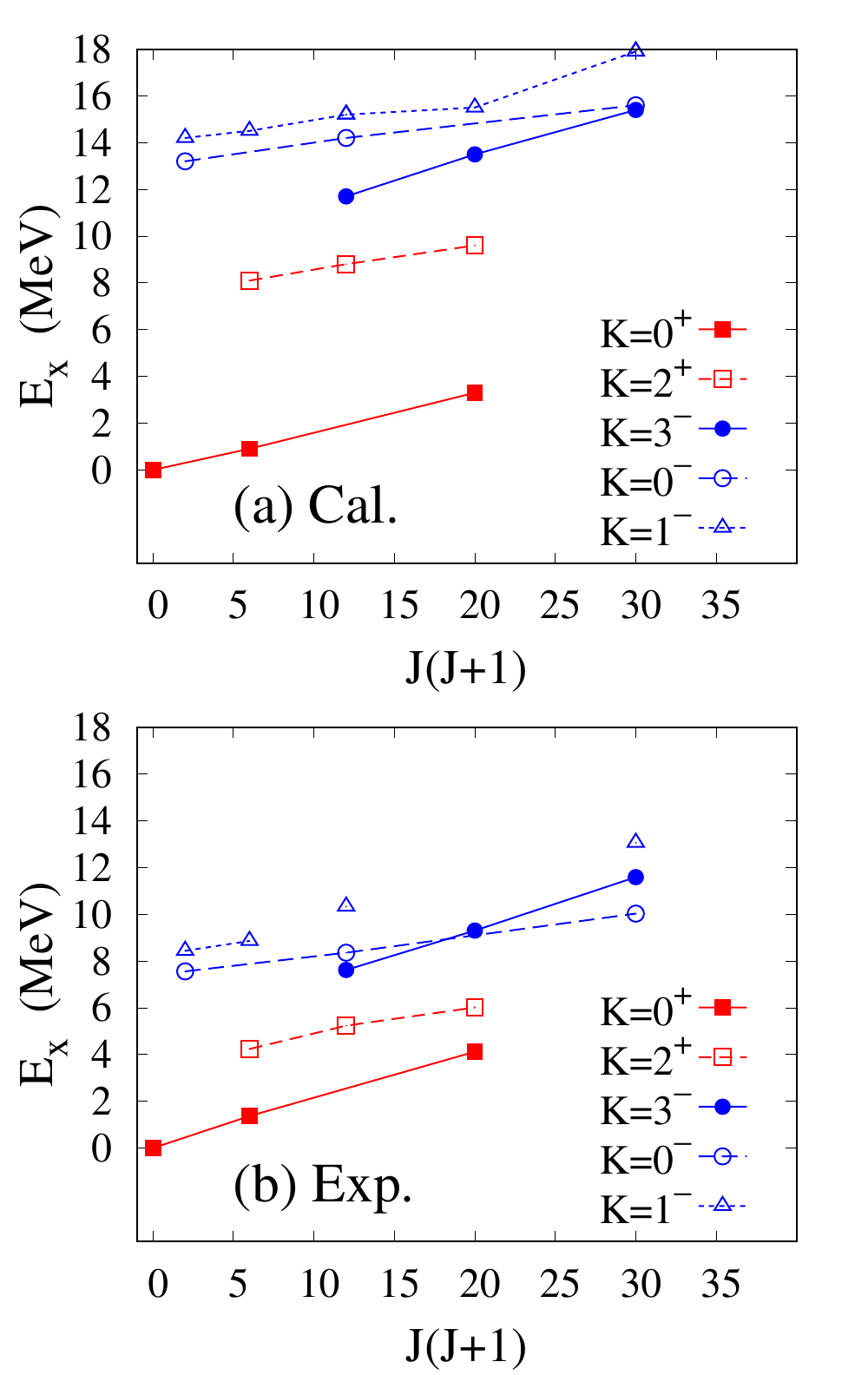}
  \caption{The energy spectra of $\Mg$. 
(a) The calculated energy levels.
(b) The experimental levels 
for the $K^\pi=2^+$ ground- and the 
$K^\pi=2^+$ side-bands from Ref.~\cite{Firestone:2007crk}, and those for the 
$K^\pi=3^-$,  $K^\pi=0^-$, and  $K^\pi=1^-$ bands assigned in Ref.~\cite{Fifield:1979gfv}.
\label{fig:spe}}
\end{figure}
%%%%%%%%%%%%%%%%%%%%%%%%%

\begin{table}[!ht]
\caption{The calculated excitation energies ($E_x$) and the root-mean-square matter radii ($R$) of $\Mg$ 
and the experimental energies for the $K^\pi=0^+$, $K^\pi=2^+$, $K^\pi=3^-$, $K^\pi=0^-$, and $K^\pi=1^-$ bands.
The calculated and experimental values of the 
electric quadrupole moment ($Q$) of the $2^+_1$ state are also shown.
The experimental energies 
are from Ref.~\cite{Firestone:2007crk}. 
For the experimental negative-parity bands, the band assignment is a tentative one 
from Ref.~\cite{Fifield:1979gfv}. 
The experimental value of the point-proton rms radius for the ground state is $R=2.941(2)$~fm 
from the charge radius data~\cite{Angeli2013}. 
 \label{tab:radii}
}
\begin{center}
\begin{tabular}{lrrrrrrrrrrccccc}
\hline
\hline
& exp     &    \multicolumn{2}{c}{AMD} \\
$J^\pi$	(band)	&	$E_x$~(MeV)	&	$E_x$~(MeV)	&	$R_m$~(fm)	\\
$0^+_1$	($K^\pi=0^+$)	&	0.00 	&	0.0 	&	3.02 	\\
$2^+_1$	($K^\pi=0^+$)	&	1.37 	&	0.9 	&	3.02 	\\
$4^+_1$	($K^\pi=0^+$)	&	4.12 	&	3.3 	&	3.01 	\\
		&		&		&		\\
$2^+_2$	($K^\pi=2^+$)	&	4.24 	&	8.1 	&	3.06 	\\
$3^+_1$	($K^\pi=2^+$)	&	5.24 	&	8.8 	&	3.06 	\\
$4^+_2$	($K^\pi=2^+$)	&	6.01 	&	9.6 	&	3.05 	\\
		&		&		&		\\
$3^-_1$	($K^\pi=3^-$)	&	7.62 	&	11.7 	&	3.02 	\\
$4^-$	($K^\pi=3^-$)	&	9.30 	&	13.5 	&	3.02 	\\
$5^-$	($K^\pi=3^-$)	&	11.59 	&	15.4 	&	3.02 	\\
		&		&		&		\\
$1^-_1$	($K^\pi=0^-$)	&	7.56 	&	13.2 	&	3.12 	\\
$3^-_2$	($K^\pi=0^-$)	&	8.36 	&	14.2 	&	3.11 	\\
$5^-$	($K^\pi=0^-$)	&	10.03 	&	15.6 	&	3.08 	\\
		&		&		&		\\
$1^-_2$	($K^\pi=1^-$)	&	8.44 	&	14.2 	&	3.10 	\\
$2^-_1$	($K^\pi=1^-$)	&	8.86 	&	14.5 	&	3.10 	\\
$3^-$	($K^\pi=1^-$)	&	10.33 	&	15.2 	&	3.10 	\\
$4^-$	($K^\pi=1^-$)	&		&	15.5 	&	3.09 	\\
$5^-$	($K^\pi=1^-$)	&	13.06 	&	17.9 	&	3.09 	\\
		&		&		&		\\
		&	$Q$~($e\textrm{fm}^2$)	&		&	$Q$~($e\textrm{fm}^2$)	\\
$2^+$	$K^\pi=0^+$	&	$-16.6(6)$	&		&	$-15.1$	\\
\hline
\hline
\end{tabular}
\end{center}
\end{table}

The AMD results of excitation energies ($E_x$) for
the $K^\pi=0^+$, $K^\pi=2^+$, $K^\pi=3^-$, $K^\pi=0^-$, and $K^\pi=1^-$ bands 
of $\Mg$ are listed 
in Table \ref{tab:radii},
together with the experimental data.
The theoretical states, $\{0^+_1,2^+_1,4^+_1 \}$, $\{2^+_2,3^+_1,4^+_2 \}$, 
 $\{3^-_1,4^-_1,5^-_1 \}$,  $\{1^-_1,3^-_2,5^-_2 \}$, and  $\{1^-_2,2^-_1,3^-_3,4^-_2,5^-_3 \}$,   
are assigned to the $K^\pi=0^+$, $K^\pi=2^+$, $K^\pi=3^-$, $K^\pi=0^-$, and  $K^\pi=1^-$ bands
based upon analysis of the $E2$-transition strengths. However, state mixing between three negative-parity 
bands is rather strong, as shown later; hence, the negative-parity band structure cannot be 
strictly defined.
For the experimental states of the negative-parity bands, 
we adopt the tentative band assignment used in Ref.~\cite{Fifield:1979gfv}. 

The calculated and experimental energy spectra are plotted in Fig.~\ref{fig:spe}. The present calculation 
well reproduces the  level spacing in each band and qualitatively describes the energy 
ordering of the positive- and negative-parity bands, but it generally overestimates the band-head 
energies of the excited bands.  
Higher-order effects beyond the present structure model may be a reason for such overestimation. 

The calculated root-mean-square radii $(R)$ of the ground and excited states 
and the electric-quadrupole moment ($Q$) of the $2^+_1$ state are listed in Table \ref{tab:radii}. 
The calculated values $R=3.02$ fm (of the ground state) and $Q=-15.1$ $e$fm$^{2}$ (of the $2^+_1$ state) 
are consistent with the observed values of $R=2.941(2)$ fm and $Q=-16.6(6)$ $e$fm$^{2}$. 
The calculation predicts slightly larger radii for 
the $K^\pi=0^-$ and $K^\pi=1^-$ bands, because they have deformations
than those of the $K^\pi=0^+$ and $K^\pi=3^-$ bands 
but the difference is small. 

\begin{table}[!ht]
\caption{$E2$-transition strengths of $\Mg$.
The experimental values $B_\textrm{exp}(E2)$ for positive- and negative-parity states are from 
Refs.~\cite{Firestone:2007crk,Keinonen:1989ltz} and Ref.~\cite{Fifield:1979gfv}, respectively.
The theoretical values $B_\textrm{th}(E2)$ obtained by the VAP calculation are listed 
together with the renormalization factors $f^\textrm{tr}$ used for the reaction 
calculations. 
The $E2$-transition strengths are in units of $e^2$fm$^{4}$.
 \label{tab:BE2}
}
\begin{center}
\begin{tabular}{ll|r|rrrrrrrrrrrccccc}
\hline
\hline
$J_i$~(band)	&	$J_f$~(band) &	exp$\quad$	&	\multicolumn{2}{c}{AMD}	&		\\
	&		&	$B_\textrm{exp}(E2)$	&	$B_\textrm{th}(E2)$	&	$f^\textrm{tr}$	\\
$2^+_1$ $(K=0)$	&	$0^+_1$ $(K=0)$	&	88.4(4.1)	&	55.4 	&	1.26 	\\
$4^+_1$ $(K=0)$	&	$2^+_1$ $(K=0)$	&	160(16)	&	72.8 	&	1.48 	\\
	&		&		&		&		\\
$3^+_1$ $(K=2)$	&	$2^+_2$ $(K=2)$	&	240(30)	&	103.1 	&		\\
$4^+_2$ $(K=2)$	&	$2^+_2$ $(K=2)$	&	77(10)	&	36.2 	&	1.46 	\\
$4^+_2$ $(K=2)$	&	$3^+_1$ $(K=2)$	&		&	73.3 	&		\\
	&		&		&		&		\\
$2^+_2$ $(K=2)$	&	$0^+_1$ $(K=0)$	&	8.0(0.8)	&	2.1 	&	1.95 	\\
$2^+_2$ $(K=2)$	&	$2^+_1$ $(K=0)$	&	12.2(0.9)	&	0.6 	&	1\footnote{no renormalization.}\\
$3^+_1$ $(K=2)$	&	$2^+_1$ $(K=0)$	&	10.3(1.2)	&	3.1 	&		\\
$4^+_2$ $(K=2)$	&	$2^+_1$ $(K=0)$	&	4.1(0.4)	&	2.9 	&	1.19 	\\
	&		&		&		&		\\
$4^-$ $(K=3)$	&	$3^-_1$ $(K=3)$	&	119(25)	&	76.9 	&		\\
$5^-$ $(K=3)$	&	$3^-_1$ $(K=3)$	&	19(6)	&	14.9 	&		\\
$5^-$ $(K=3)$	&	$4^-$ $(K=3)$	&	152(45)	&	48.7 	&		\\
	&		&		&		&		\\
$3^-_2$ $(K=0)$	&	$1^-_1$ $(K=0)$	&		&	67.9 	&		\\
$5^-$ $(K=0)$	&	$3^-_2$ $(K=0)$	&	82(27)	&	81.4 	&		\\
	&		&		&		&		\\
$2^-_1$ $(K=1)$	&	$1^-_2$ $(K=1)$	&		&	132 	&		\\
$3^-$ $(K=1)$	&	$1^-_2$ $(K=1)$	&		&	45.8 	&		\\
$3^-$ $(K=1)$	&	$2^-_1$ $(K=1)$	&		&	32.9 	&		\\
$4^-$ $(K=1)$	&	$2^-_1$ $(K=1)$	&		&	101.5 	&		\\
$4^-$ $(K=1)$	&	$3^-$ $(K=1)$	&		&	27.6 	&		\\
$5^-$ $(K=1)$	&	$4^-$ $(K=1)$	&		&	5.6 	&		\\
$5^-$ $(K=1)$	&	$3^-$ $(K=1)$	&	90(16)	&	91.5 	&		\\
	&		&		&		&		\\
$3^-_2$ $(K=0)$	&	$1^-_2$ $(K=1)$	&		&	17.7 	&		\\
$3^-_2$ $(K=0)$	&	$2^-_1$ $(K=1)$	&		&	34.1 	&		\\
$5^-$ $(K=0)$	&	$4^-$ $(K=1)$	&		&	19.9 	&		\\
$3^-$ $(K=1)$	&	$1^-_1$ $(K=0)$	&		&	36.4 	&		\\
$4^-$ $(K=1)$	&	$3^-_2$ $(K=0)$	&		&	30.3 	&		\\
$5^-$ $(K=1)$	&	$3^-_2$ $(K=0)$	&	2.7(0.7)	&	20.1 	&		\\
$5^-$ $(K=0)$	&	$4^-$ $(K=3)$	&		&	10.3 	&		\\
$5^-$ $(K=3)$	&	$3^-_2$ $(K=0)$	&		&	14.4 	&		\\
\hline
\hline
\end{tabular}
\end{center}
\end{table}

\begin{table*}[ht]
\caption{$E\lambda(C\lambda)$ and isoscalar IS$\lambda$
transition strengths to the $0^+_1$ state for
the $J^\pi$~($J=\lambda$) states of $\Mg$.
For the experimental values, 
$B(E\lambda;\lambda^{\pi} \to 0^+_1)$ from the $\gamma$-decay data \cite{Firestone:2007crk,KIBEDI:2002tqu}, 
$B(C\lambda;\lambda^{\pi} \to 0^+_1)$ from the $(e,e')$ data \cite{Zarek:1978cvz,Zarek:1984fm,Johnston_1974}, 
$B(\textrm{IS}\lambda;\lambda^{\pi} \to 0^+_1)/4$ from the 
$(\alpha,\alpha')$ data \cite{VanDerBorg:1981qiu}, and $B(C\lambda;\lambda^{\pi} \to 0^+_1)$ from the $(\pi,\pi')$ 
data~\cite{Blanpied:1990vd} are listed.
For theoretical values, the original values $B_\textrm{th}(IS\lambda)/4$ before renormalization and 
the renormalized values $(f^\textrm{tr})^2B_\textrm{th}(IS\lambda)/4$ used for the reaction calculations
are shown together with the adopted renormalization factors $f^\textrm{tr}$. 
Transition strengths are in units of $e^2$fm$^{2\lambda}$ for the $\lambda=2,3$ and 4 transitions, and  
$e^2$fm$^{6}$ for the isoscalar dipole~(IS1) transitions.
\label{tab:BEL}
}
\begin{center}
\begin{tabular}{lrrrrrrrrrrrrrccccc}
\hline
\hline
	&	$\gamma$-decays	&	$(e,e')$	&	$(e,e')$	&	$(\alpha,\alpha')$	&	$(\pi,\pi')$	&	\multicolumn{3}{c}{AMD}	\\
$J^\pi$~(band)	&	$B(E\lambda)$	&	$B(C\lambda)$	&	$B(C\lambda)$	&	$B(\textrm{IS}\lambda)/4$	&	$B(C\lambda)$	
&	\multicolumn{2}{c}{$B(\textrm{IS}\lambda)/4$}	&	$f^\textrm{tr}$	\\
	&	\cite{Firestone:2007crk,KIBEDI:2002tqu}	&	\cite{Zarek:1978cvz,Zarek:1984fm}	&	\cite{Johnston_1974}	&	
\cite{VanDerBorg:1981qiu}	&	\cite{Blanpied:1990vd}	&	original	&	normalized	&		\\
$2^+_1$ $(K=0)$	&	88.4(4.1)	&	90.6(7.0)	&	105(5)	&	84 	&	108	&	54 	&	86 	&	1.26	\footnote{$f^\textrm{tr}$ determined to fit $B(E\lambda)$ from $\gamma$-decays.}	\\
$2^+_2$ $(K=0)$	&	8.0(0.8)	&	5.48(0.60)	&	5.26(1.2)	&	14 	&	6.7	&	2.0 	&	7.7 	&	1.95	$^a$	\\
$4^+_1$ $(K=0)$	&		&	200(30)	&		&	1200 	&		&	1.1 	&	1.1 	&	1.0 	\footnote{No renormalization.}	\\
$4^+_2$ $(K=0)$	&		&	4800(600)	&	4700(1100)	&	4700 	&	2900	&	1740 	&	4800 	&	1.66	\footnote{$f^\textrm{tr}$ determined to fit $B(C\lambda)$ from $(e,e')$ data~\cite{Zarek:1978cvz,Zarek:1984fm}.}	\\
	&		&		&		&		&		&		&		&			\\
$3^-_1$ $(K=3)$	&	221(44)	&	80 	&	190(30)	&	190 	&	136 	&	28 	&	80 	&	1.68	$^c$	\\
$3^-_2$ $(K=0)$	&		&	226 	&	290(30)	&	280 	&	226 	&	89 	&	226 	&	1.59 	$^c$	\\
$3^-$ $(K=1)$	&		&		&		&		&		&	0.1 	&	0.1 	&	1.0 	$^b$	\\
$1^-_1$ $(K=0)$	&		&		&		&		&		&	3.1 	&	19.3 	&	2.5	\footnote{$f^\textrm{tr}$ determined to fit the charge-form factors from $(e,e')$ data~\cite{Zarek:1978cvz,Zarek:1984fm}.}	\\
$1^-_2$ $(K=1)$	&		&		&		&		&		&	4.3 	&	17.2 	&	2.0 	$^d$	\\
\hline
\end{tabular}
\end{center}
\end{table*}

In Table~\ref{tab:BE2}, 
the calculated $E2$-transition strengths are compared with the experimental data. 
For the negative-parity states,
the assignment of the $K^\pi=3^-$, $K^\pi=0^-$, and $K^\pi=1^-$ bands is done tentatively 
for calculated states having remarkably strong $E2$ transitions. However, strong $E2$ transitions 
are also obtained for inter-band transitions, in particular, between the $K^\pi=0^-$ and $K^\pi=1^-$ bands,  
and indicate strong band mixing, 
The present AMD calculation qualitatively describes the experimental $E2$-transition strengths, 
but the quantitative agreement with the data is unsatisfactory.
The theoretical strengths $B_\textrm{th}(E2)$ tend to underestimate the experimental data
$B_\textrm{exp}(E2)$,
possibly because the present AMD calculation is a simple version 
based on the single-Slater description of spherical Gaussians and 
may be insufficient to describe the large collectivity of deformations in $\Mg$.
To use the transition densities in the MCC calculations, we introduce the renormalization factors 
$f^\textrm{tr}=(B_\textrm{exp}/B_\textrm{th})^{1/2}$ to fit the observed values $B_\textrm{exp}(E2)$,
as mentioned previously. The $f^\textrm{tr}$ values adopted in the present MCC calculation
are given in Table~\ref{tab:BE2}. 
The factors $f^\textrm{tr}=1.19$--1.95 are needed to fit the $B_\textrm{exp}(E2)$ values. 
It should be noted that 
bare charges of nucleons are adopted in the AMD framework, 
unlike the shell models in which the effective charges of protons and neutrons are usually required.
If we introduce the effective charges,
the values of $f^\textrm{tr}=1.19$--1.95 obtained in the present AMD result correspond to the enhancement 
$\delta=0.1$--0.5 for the effective charges 
$e^\textrm{eff}_p=1+\delta$ and $e^\textrm{eff}_n=\delta$ of the protons and neutrons, which are comparable
to or even smaller than standard values of shell models.

In Table~\ref{tab:BEL}, 
the results for the inelastic transition strengths
from the ground state are listed in comparison with the experimental values
measured by $\gamma$-decays and evaluated by electron, $\alpha$, and pion scattering.
For the $E2$, $E3$, and $E4$ transitions,  
the calculated values 
$B(\textrm{IS}\lambda)/4$ of the isoscalar component are compared with the 
experimental data for $B(E\lambda)$, $B(C\lambda)$, and $B(\textrm{IS}\lambda)/4$. 
For the dipole transitions, the calculated values of the isoscalar dipole (IS1) transition strengths 
are shown in the table.
For use in the MCC calculations,
the renomalization factors $f^\textrm{tr}$ for the $0^+_1\to 3^-_1$, 
$0^+_1\to 3^-_2$, and $0^+_1\to 4^-_2$ transitions  
are determined to fit the $B(C\lambda)$ values that were evaluated 
from the $(e,e')$ data. For the $\textrm{IS}1$ transitions, $f^\textrm{tr}$  
are determined to fit the charge-form factors measured by the $(e,e')$ experiments.
As a result of this fitting, the renormalization factors
for the $4^+_2$($K^\pi=2^+$), 
$3^-_1$($K^\pi=3^-$), $3^-_2$($K^\pi=0^-$), $1^-_1$($K^\pi=0^-$), and $1^-_2$($K^\pi=1^-$) states 
are obtained in the range of $f^\textrm{tr}=1.59$--2.5, which again means that 
the collectivity of these excited states is somewhat underestimated by the present AMD calculation. 

For the $4^+_1$($K^\pi=0^+$) state, a remarkably weak $\lambda=4$ transition
wad observed in 
the $(e,e')$ experiment~\cite{Zarek:1978cvz}. The $0^+_1\to 4^+_1$ transition strength is 
more than one order smaller than that for the $0^+_1\to 4^+_2$ transition meaning that 
the $\lambda=4$ strength from the ground state is dominantly concentrated not in the 
$4^+_1$($K^\pi=0^+$) state but rather in the $4^+_2$($K^\pi=2^+$) state.
The calculation describes this trend of weak $\lambda=4$ transition in the $K^\pi=0^+$
ground-band, but quantitatively it is too weak compared with the observed data and hence 
$f^\textrm{tr}=1$ (no renormaliation) is adopted for this transition in the MCC calculation. 
The leading feature of the $4^+_1$($K^\pi=0^+$) state is the strong in-band $E2$ transition to the 
$2^+_1$($K^\pi=0^+$) state, whereas the $E4$ transition is a higher-order effect. 
We can say that the present calculation qualitatively reproduces the leading feature of the $4^+_1$ state 
but fails to describe the higher-order effect.

\subsection{Intrinsic structure of $\Mg$: AMD results}

To discuss the intrinsic structure of the ground and excited bands, 
we analyze the single-Slater AMD wave functions for the dominant configurations
of the band-head states, 
which are obtained by VAP.
Such a simple analysis is useful for obtaining an intuitive understanding of the leading features 
though the final wave functions are affected by state mixing and in-band structure change.

\begin{figure}[!h]
\includegraphics[width=9 cm]{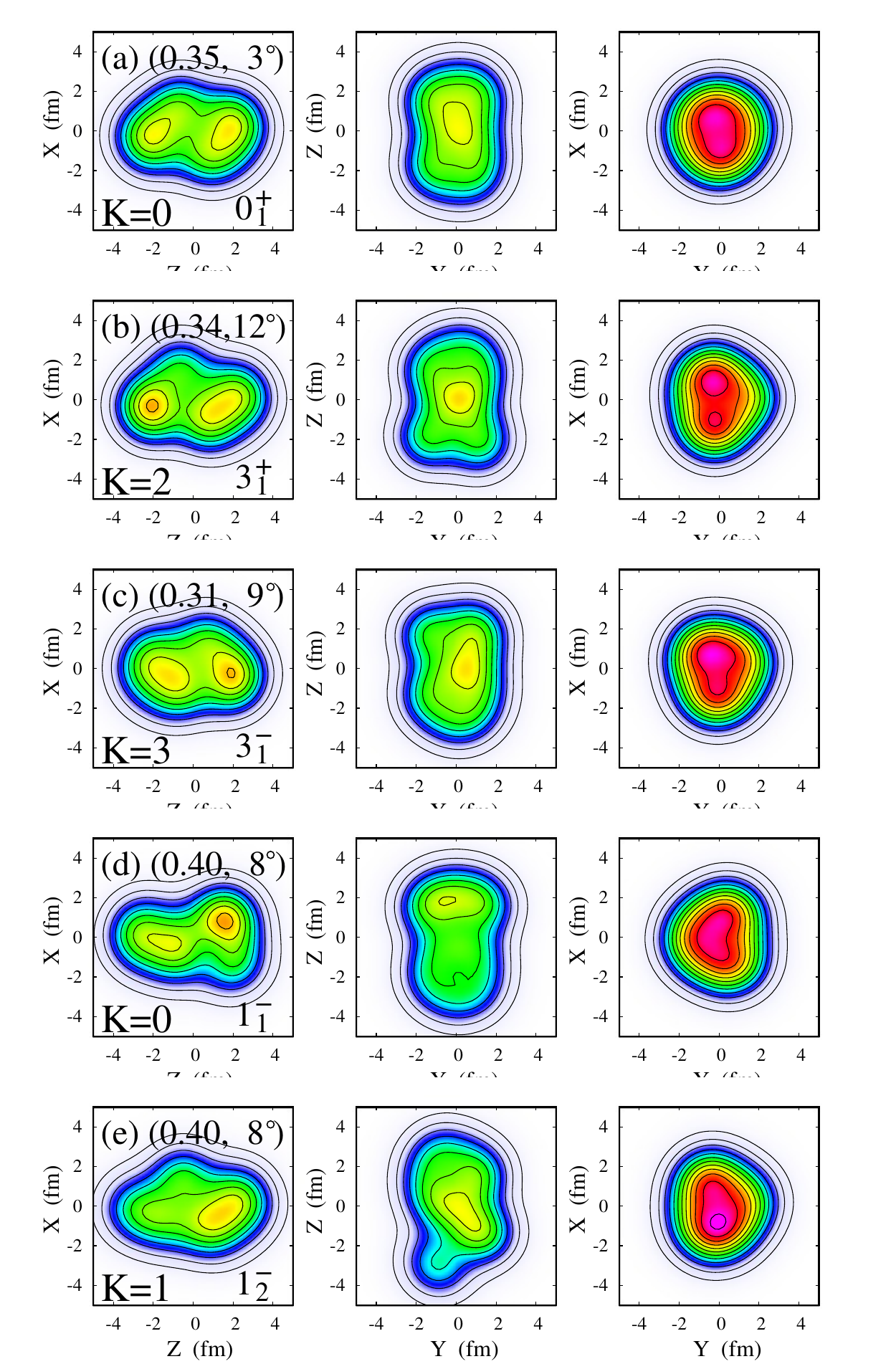}
  \caption{Density distribution of intrinsic wave functions prior tp the parity and total-angular-momentum projections for the $0^+_1$($K^\pi=0^+$), $3^+_1$($K^\pi=2^+$), $3^-_1$($K^\pi=3^-$), 
$1^-_1$($K^\pi=0^-$), and $1^-_1$($K^\pi=1^-$) states, as obtained by AMD.
The integrated density projected onto the $X$-$Z$, 
$Y$-$Z$, and $Y$-$X$ planes is plotted in the left, middle, and right panels, respectively, 
by contours with the interval of 0.1 fm$^{-2}$ interval.
For each state, the axes are chosen to be the principal axes of intrinsic deformation as 
$\langle ZZ\rangle\ge \langle YY\rangle\ge \langle XX\rangle$ and 
 $\langle XY\rangle=\langle YZ\rangle=\langle ZX\rangle=0$.
The deformation parameters $(\beta,\gamma)$ calculated from the expectation values, 
$\langle ZZ\rangle$,  $\langle YY\rangle$, and $\langle XX\rangle$, are shown in the left panels.
  \label{fig:dense-cont}}
\end{figure}

Figure~\ref{fig:dense-cont} shows the 
density distribution of the intrinsic wave functions before the parity and total-angular-momentum projections for the $0^+_1$($K^\pi=0^+$), $3^+_1$($K^\pi=2^+$), $3^-_1$($K^\pi=3^-$), 
$1^-_1$($K^\pi=0^-$), and $1^-_1$($K^\pi=1^-$) states. The quadrupole-deformation 
parameters $\beta,\gamma$, which are calculated from the expectation values of 
$\langle ZZ\rangle$,  $\langle YY\rangle$, and $\langle XX\rangle$ for the intrinsic wave functions, 
are also shown. The $K^\pi=0^+$ band has an approximately prolate deformation
with a $^{12}$C+$^{12}$C clustering feature. 
The deformation is $\beta=0.35$ in the $0^+_1$ state, and gradually decreases to $\beta=0.33$
and 0.29 in the $2^+_1$($K^\pi=0^+$) and $4^+_1$($K^\pi=0^+$) states, respectively.  
%because of the anti-stretching effect.
 
The excited bands have triaxial deformations with $^{12}\textrm{C}+3\alpha$-like cluster structures.
In particular, the $K^\pi=2^+$ band has the largest triaxiality as $\gamma=12^\circ$,
because of the $2\alpha$ clustering around the $^{12}\textrm{C}+\alpha$ core part
as shown in Fig.~\ref{fig:dense-cont}(b). 
The $\beta$ deformation of the $K^\pi=2^+$ band is approximately the same as
that of the ground state, meaning that this band can be understood as the $K^\pi=2^+$ side-band 
of the $K^\pi=0^+$ ground-band. 
For the negative-parity bands, the $K^\pi=3^-$ band has almost the same $\beta$ deformation
as the ground state, whereas the $K^\pi=0^-$ and $K^\pi=1^-$ bands have larger deformations
as $\beta=0.40$. In the intrinsic densities for the $K^\pi=0^-$ and $K^\pi=1^-$ bands 
shown in Figs.~\ref{fig:dense-cont}(d) and (e), 
one can see that the reflection symmetry for $Z\leftrightarrow -Z$ in the $Z$ direction is broken 
in the $K^\pi=0^-$ band owing to the asymmetric structure of $^{12}\textrm{C}+^{12}\textrm{C}$ clustering, 
whereas the reflection symmetry for $Y\leftrightarrow -Y$ in the $Y$ direction 
is broken in the $K^\pi=1^-$ band.
These symmetry-broken shapes produce two types of negative-parity excitations with
quanta of $K=0$ and $K=1$. 
This result for the $K^\pi=0^-$ and $K^\pi=1^-$ bands is similar to that for the deformed AMD result
for low-lying $1^-$ states, as found in Ref.~\cite{Chiba:2019dap}, although that 
calculation yielded the reverse ordering of the $K^\pi=0^-$ and $K^\pi=1^-$ energies. 

\begin{figure}[!h]
\includegraphics[width=9 cm]{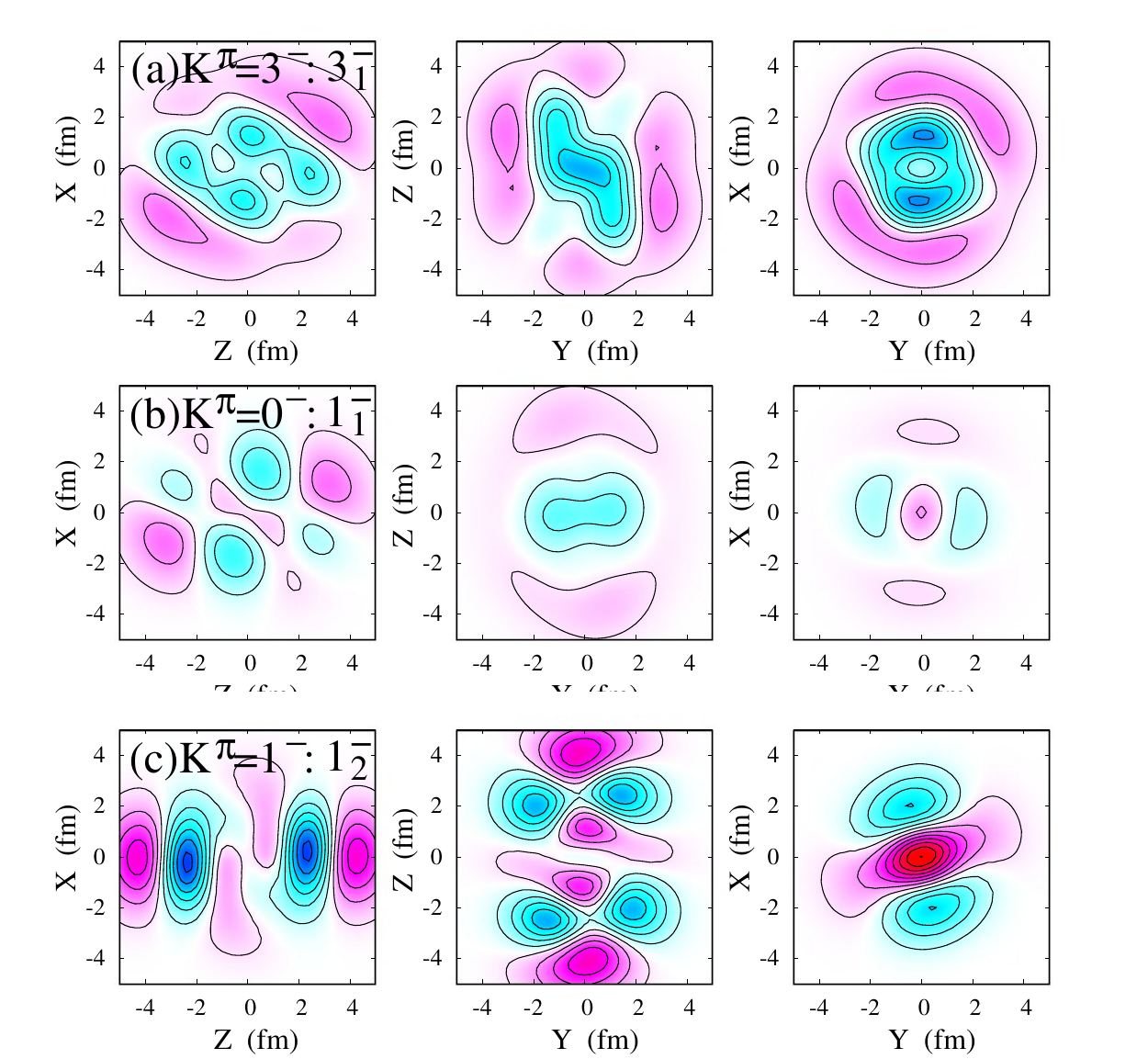}
\caption{Density difference between the positive-parity and negative-parity components 
in the intrinsic states of the band-head states; 
(a) $3^-_1(K^\pi=3^-)$, (b) $1^-_1(K^\pi=0^-)$, and 
(c) $1^-_2(K^\pi=1^-)$.
For each state, 
The intrinsic density of the positive-parity component is subtracted from that of 
the negative-parity component, and  
the difference of the integrated densities is projected onto the $Z$-$X$, 
$Y$-$Z$, and $Y$-$X$ planes, as shown in the left, middle, and right of the figure, respectively.
The intrinsic axes are chosen to for consistency with 
Fig.~\ref{fig:dense-cont}.
The contour interval is 0.003 fm$^{-2}$, and the red (blue) color map indicate  
positive (negative) values.
\label{fig:dense-cont-np}}
\end{figure}

The negative-parity bands are built on different kinds of excitation modes;
these excitations contain large-amplitude dynamics and cluster correlations
beyond the single-particle excitations on the ground state in the mean-field picture.
However, with the help of single-particle analyses of the present AMD configurations, 
we can associate the leading aspects of the $K^\pi=3^-$ and $K^\pi=1^-$ bands of 
$\Mg$ with $1p$-$1h$ excitations in the deformed state. In Fig.~\ref{fig:dense-cont-np}, we show the 
density difference between the positive- and negative-parity components
in each band-head state of the $3^-_1$($K^\pi=3^-$), 
$1^-_1$($K^\pi=0^-$), and $1^-_1$($K^\pi=1^-$) states. 
For each intrinsic wave function, 
the positive- and negative-parity
components are normalized as 
$|P^\pi \Phi_\textrm{AMD}\rangle/\sqrt{\langle 
P^\pi\Phi_\textrm{AMD}|P^\pi \Phi_\textrm{AMD}\rangle}$, 
and  the positive-parity density is subtracted from the negative-parity density. 
This density difference may reflect the $1p$-$1h$ feature, by which 
positive- and negative-sign contributions correspond to 
particle and hole densities, respectively. 
In the density difference for the  $K^\pi=3^-$ band 
(Fig.~\ref{fig:dense-cont-np}(a)), the hole contribution is remarkable in the inner region
and indicates significant contributions by single-particle excitations from the 
$p$-shell to the $sd$-shell; this is associated with the  
$K^\pi=3^-$ excitation of the $^{16}\textrm{O}$ core due to $^{12}\textrm{C}+\alpha$ clustering.
The $1^-_2(K^\pi=1^-)$ state clearly shows the $(sd)^{-1} (fp)$ feature shown in 
Fig.~\ref{fig:dense-cont-np}(c), which is 
predominantly interpreted as the $(0,1,1)^{-1}(0,0,3)$ configuration in terms of 
the single-particle description $(n_x,n_y,n_z)$ of the three-dimensional
oscillator quanta. 
This configuration of the $K^\pi=1^-$ band corresponds to the toroidal dipole excitation 
of the deformed state, which was obtained in Refs.~\cite{Nesterenko:2017rcc,Nesterenko:2019dnt,Chiba:2019dap} 
not as the second $1^-$ state but as the lowest $1^-$ state. 
For the $1^-_1(K^\pi=0^-)$ state, 
the single-particle aspects are unclear, but one can see a signal from the mixture of the 
$(0,1,1)^{-1}(0,1,2)$ and $(0,0,2)^{-1}(0,0,3)$ configurations, as caused by 
the $K^\pi=0^-$ excitation of the cluster mode (see the left panel of Fig.~\ref{fig:dense-cont-np}(b)).

%%%%%%%%%%%%%%%%%%%%%%%%%%%%%%
%%%%%%%%%%%%%%%%%%%%%%%%%

%%%%%%%%%%%%%%%%%%%%%%%%%%%%%%
%%%%%%%%%%%%%%%%%%%%%%%%%

\subsection{Diagonal and transition densities and charge-form factors}

%%%%%%%%%%%%%%%%%%%%%%%%%%%%%%
\begin{figure}[!h]
\includegraphics[width=6 cm]{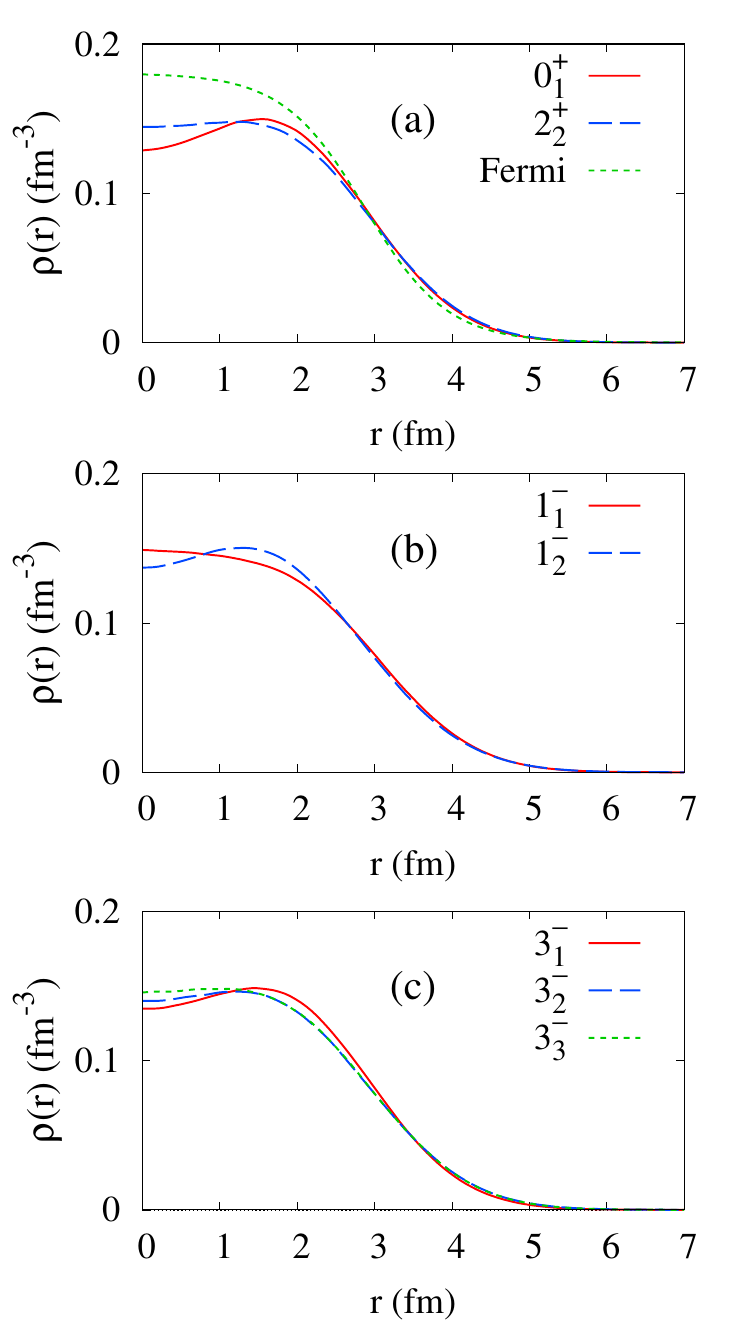}
  \caption{The matter densities of $\Mg$. 
The Fermi density $\rho^{}_\textrm{Fermi}(r)=\rho _0 [1+\exp(\frac{r-c}{t/4.4}) ]^{-1}$ 
with $c=2.876$~fm and $t=2.333$~fm is also shown in panel (a).
  \label{fig:dense}}
\end{figure}
%%%%%%%%%%%%%%%%%%%%%%%%%

In Fig.~\ref{fig:dense}, we show the matter densities of the ground 
and excited states. The $K^\pi=0^-$ and $K^\pi=1^-$ bands have slightly 
broader density tails 
than do the $K^\pi=0^+$ and $K^\pi=3^-$ bands
due to larger deformations; however, the difference in the diagonal density is small. 

\begin{figure}[!h]
\includegraphics[width=9 cm]{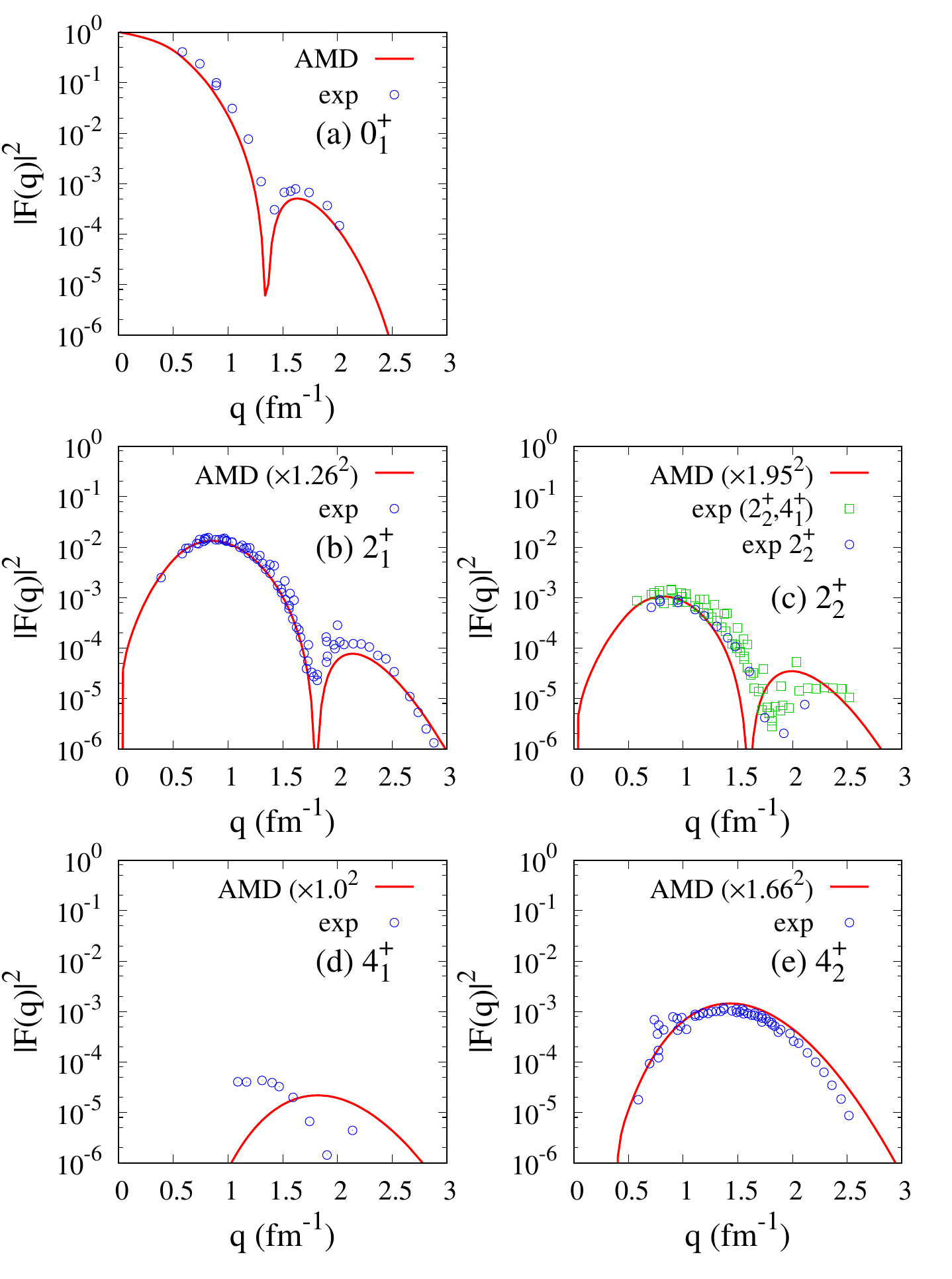}
  \caption{Square of the charge-form factors of the elastic and inelastic processes 
for the positive-parity states of $\Mg$.
For the calculated result, the square of the renormalized form factors $F(q)$ 
multiplied by the $f^\textrm{tr}$ values in Table~\ref{tab:BEL} are plotted.
The experimental data were measured by electron scattering \cite{Horikawa:1971oau,Nakada:1972,Zarek:1978cvz,Li:1974vj,Johnston_1974}.
In Refs.~\cite{Horikawa:1971oau,Nakada:1972,Li:1974vj}
for the $2^+_2$(4.24 MeV) state, 
the $4^-_1$(4.12 MeV) contributions were not separated. 
  \label{fig:form-pp}}
\end{figure}

\begin{figure}[!h]
\includegraphics[width=9 cm]{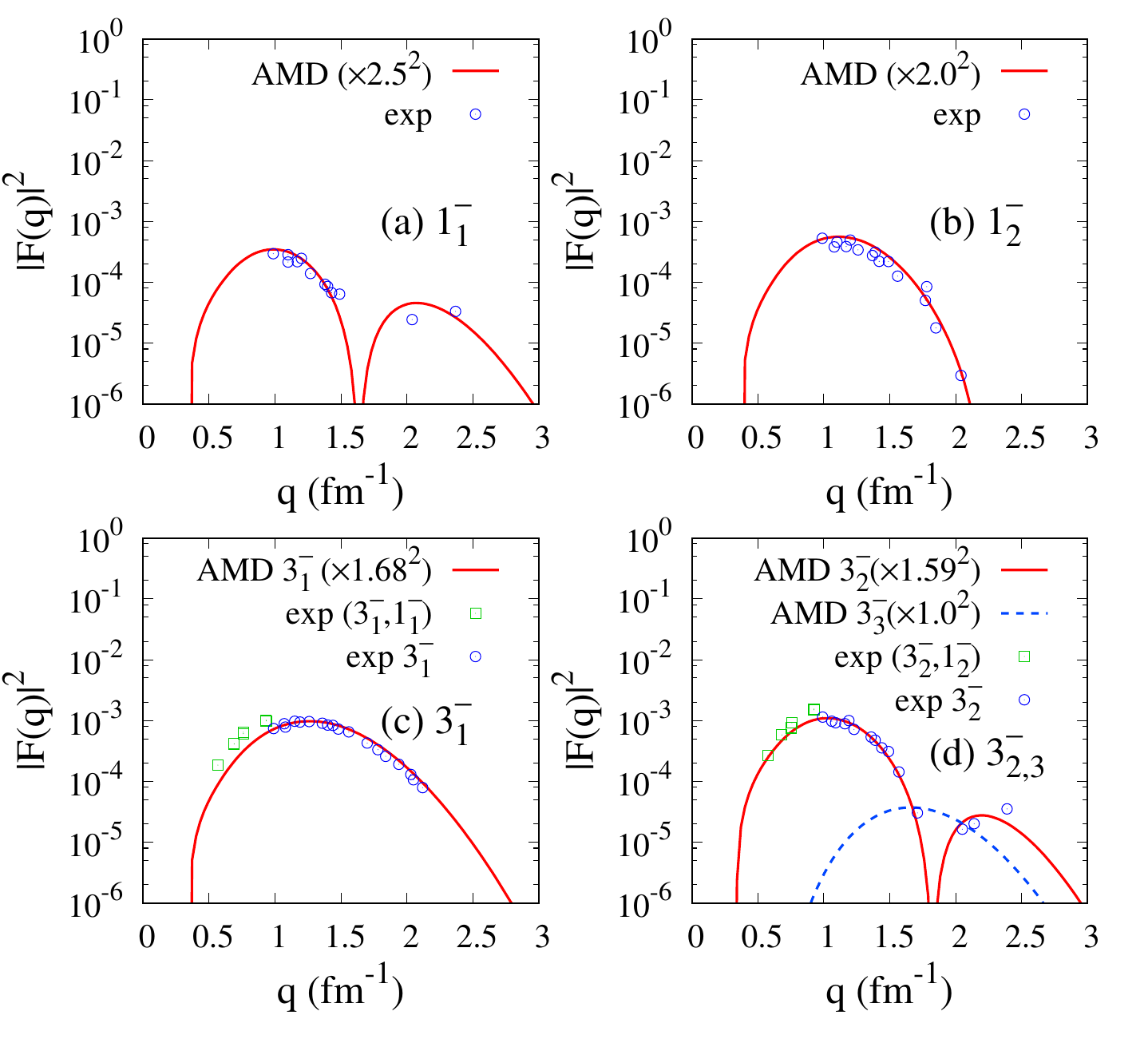}
  \caption{Same as Fig.~\ref{fig:form-pp} but for the negative-parity states.
The experimental data are taken from Refs.~\cite{Johnston_1974,Zarek:1984fm}.
In the data from Ref.~\cite{Johnston_1974} for the $3^-_1$(7.62 MeV) and $3^-_2$(8.36 MeV) states, 
the $1^-_1$(7.56 MeV) and $1^-_2$(8.44 MeV) contributions were not separated. 
  \label{fig:form-np}}
\end{figure}

Let us discuss the properties of the form factors and transition densities. 
For the calculated results, we show the renormalized 
form factors and transition densities, which are multiplied by the factors
$f^\textrm{tr}$ given in Table~\ref{tab:BEL}.  
The renormalized inelastic charge-form factors for the positive- and 
negative-parity states
are compared with the experimental data measured by $(e,e')$ 
in Figs.~\ref{fig:form-pp} and \ref{fig:form-np}, respectively. The calculated and experimental 
elastic charge-form factors are also shown in Fig.~\ref{fig:form-pp}.

The calculated form factors reproduce the state-dependent shapes of the observed form factors
(except for the $4^+_1$ state) 
and after the renormalization, they agree well with the 
experimental data. In particular, the calculation 
successfully reproduces the two-peak structure of 
the $1^-_1$ and $3^-_2$ form factors for transitions 
to the $K^\pi=0^-$ band. Hence, the strong-state dependence of the charge-form factors observed 
in the $1^-_1$, $1^-_2$, $3^-_1$, and $3^-_2$ states is well described. 
From these good agreements with the data, 
the present assignment for the calculated states to the observed states can be said to be reasonable. 
The form factors predicted for the $3^-_3$ state are remarkably small and their shape is 
inconsistent with neither the $3^-_1$ nor $3^-_2$ states 
meaning that this state has a quite different character from the lowest two $3^-$ states.  
The form factors observed for the $4^+_1$ state are much smaller than those 
for the $4^+_2$ state~\cite{Zarek:1978cvz}. The present AMD calculation yields small form factors for the 
$4^+_1$ state, which are comparable to the experimental data; however, the shape of the form factors is inconsistent with that is observed. 

\begin{figure}[!h]
\includegraphics[width=6 cm]{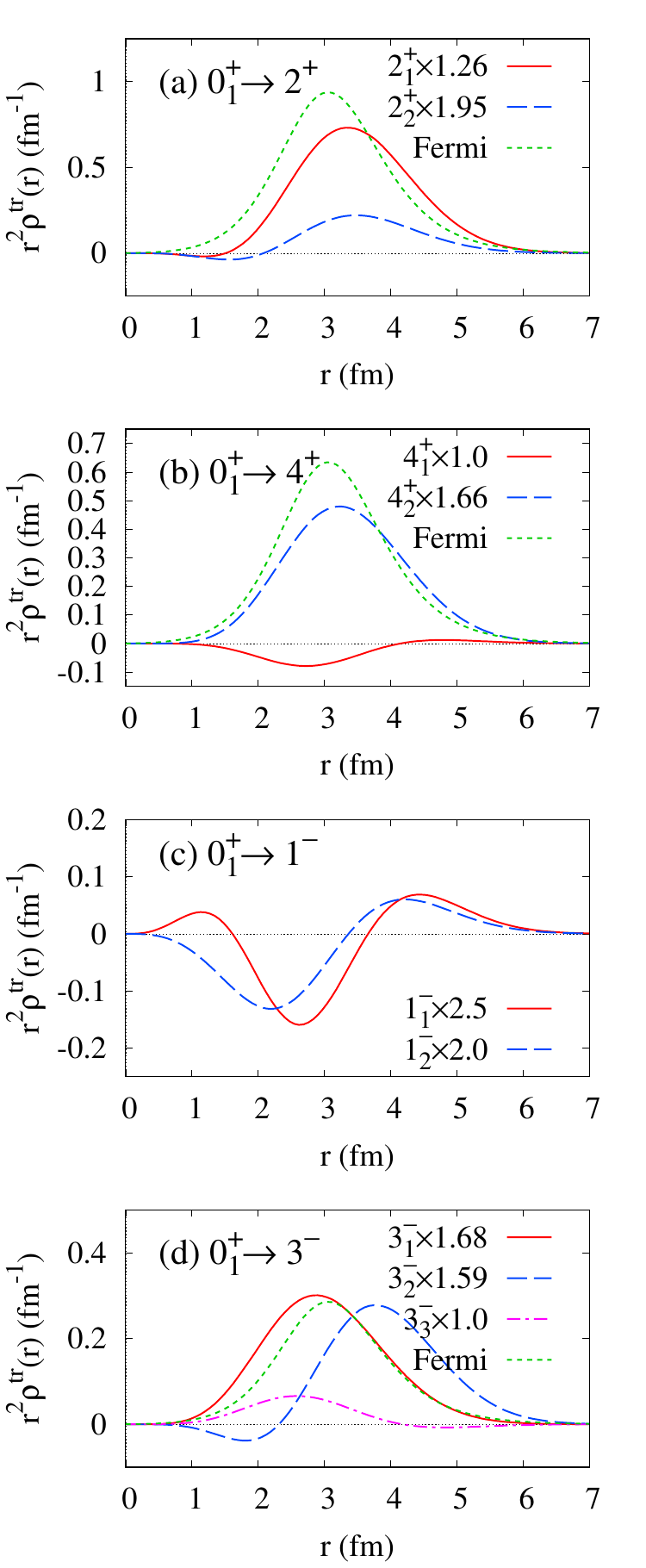}
  \caption{The isoscalar components (average of the proton and neutron components) of 
the transition densities of $\Mg$.
The renormalized  transition densities 
multiplied by the $f^\textrm{tr}$ values in Table~\ref{tab:BEL}
are plotted. 
For transitions to the $2^+$, $4^+$, and $3^-$ states, 
the collective-model-transition density of the Fermi-type Tassie form 
$\rho^\textrm{tr}_\textrm{Tassie}(r)\propto r^{\lambda-1}\partial \rho^{}_\textrm{Fermi}(r)/\partial r$
is also shown for comparison. $\rho^\textrm{tr}_\textrm{Tassie}(r)$ is normalized to fit the 
$E\lambda$-transition strengths from the $0^+_1$ state to the $2^+_1$, $4^+_2$, and $3^-_1$ states.
  \label{fig:trans}}
\end{figure}

The renormalized transition densities are shown in Fig.~\ref{fig:trans}.
The $1^-_1$ and $3^-_2$ states in the $K^\pi=0^-$ band show characteristics quite 
different from those of normal IS1 and $E3$ transitions.
As discussed previously, the calculated form factors for these states have 
narrow two-peak structures, which correspond
to the transition density broadly distributed in the outer region
with an extra node in the inner region (see Figs.~\ref{fig:trans}(c) and (d)). 
This unusual behavior is caused by the 
$K^\pi=0^-$ excitation of the asymmetric cluster structure, which involves 
radial excitations of higher-nodal orbits, including the $(1s)^{-1}(1p)$ configuration in the higher shells.
Such  nodal behavior cannot be described by collective models and indicates the
importance of a microscopic description of the inelastic transitions in the 
$K^\pi=0^-$ band. To clarify this, we also show the collective-model-transition density with 
the Fermi-type Tassie form given by the derivative form $\rho^\textrm{tr}_\textrm{Tassie}(r)\propto r^{\lambda-1}\partial \rho^{}_\textrm{Fermi}(r)/\partial r$
of the Fermi density 
\begin{align}
\rho^{}_\textrm{Fermi}(r)=\frac{\rho _0}{1+\exp(\frac{r-c}{t/4.4})},
\end{align}
where the parameters $c$ and $t$ are set to be $c=2.876$~fm and $t=2.333$~fm, respectively,
which have been adjusted in Ref.~\cite{Johnston_1974} to fit 
the elastic form factors measured by electron scattering. Note that 
the value of $2.876=\sqrt{2.985^2-0.8^2}$ fm for the point-nucleon transition density
is derived from the original 
value of $c=2.985$ fm from Ref.~\cite{Johnston_1974} for the charge-form factor
considering the proton charge-form factor.
In Fig.~\ref{fig:trans}, we compare the collective-model-transition density
for the $2^+$, $4^+$, and $3^-$ states.  
This collective-model-transition density yields 
a single-peak structure at the nuclear surface, 
which seems reasonable for the $0^+\to 2^+_1$, $0^+\to 4^+_2$, and  $0^+\to 3^-_1$
transitions but fails to describe the peak position of the $0^+\to 3^-_2$ transition in the outer region. 
%For the $3^-_1(K^\pi= 3^-_1)$ state, the calculated transition density shows peak amplitude  
%in the relatively inner region because of the $p^{-1}(sd)^1$ contribution compared with the collective-model %expectation.

%%%%%%%%%%%%%%%%%%%%%%%%%%%%%%
%%%%%%%%%%%%%%%%%%%%%%%%%

%%%%%%%%%%%%%%%%%%%%%%%%%%%%%%
%%%%%%%%%%%%%%%%%%%%%%%%%

%%%%%%%%%%%%%%%%%%%%%%%%%%%%%%
%%%%%%%%%%%%%%%%%%%%%%%%%

%%%%%%%%%%%%%%%%%%%%%%%%%%%%%%
\begin{figure*}[!h]
\includegraphics[width=18. cm]{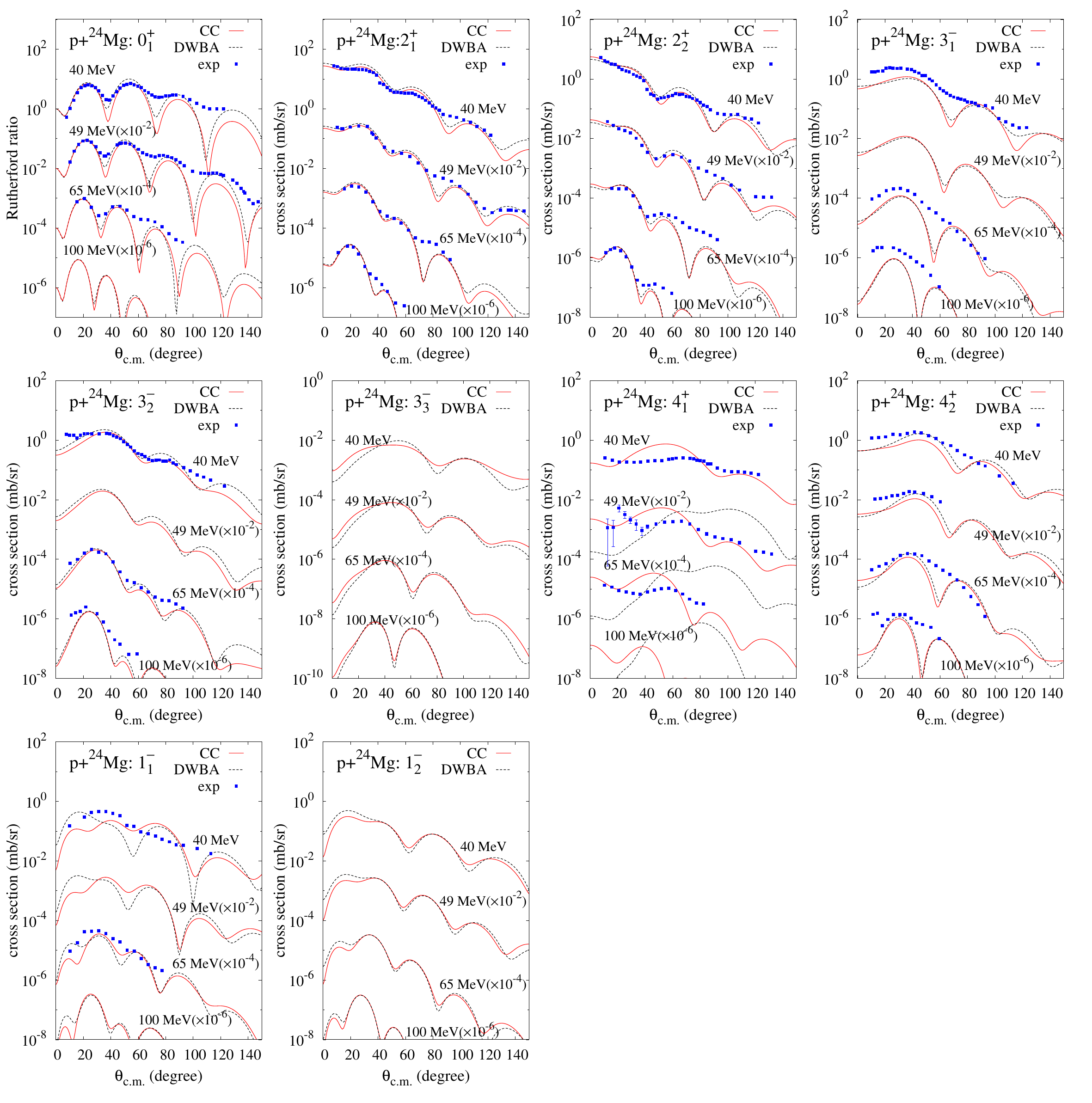}
  \caption{
Cross sections of proton  scattering off $\Mg$ 
at incident energies of $E_p=40$, 49, 65, and 100 MeV, as calculated with 
MCC+AMD (solid lines with label ``CC'') and DWBA (dotted lines with label ``DWBA'').
Experimental data are cross sections at 
$E_p=40$~MeV~\cite{Zwieglinski:1978zza,Zwieglinski:1978zz},  49~MeV~\cite{Rush:1967zwr},
65~MeV~\cite{Kato:1985zz,Otuka:2014wzu}, and 100~MeV~\cite{Horowitz:1969eso,Otuka:2014wzu}.
  \label{fig:cross-mg24p}}
\end{figure*}
%%%%%%%%%%%%%%%%%%%%%%%%%

%%%%%%%%%%%%%%%%%%%%%%%%%%%%%%
\begin{figure*}[!h]
\includegraphics[width=18. cm]{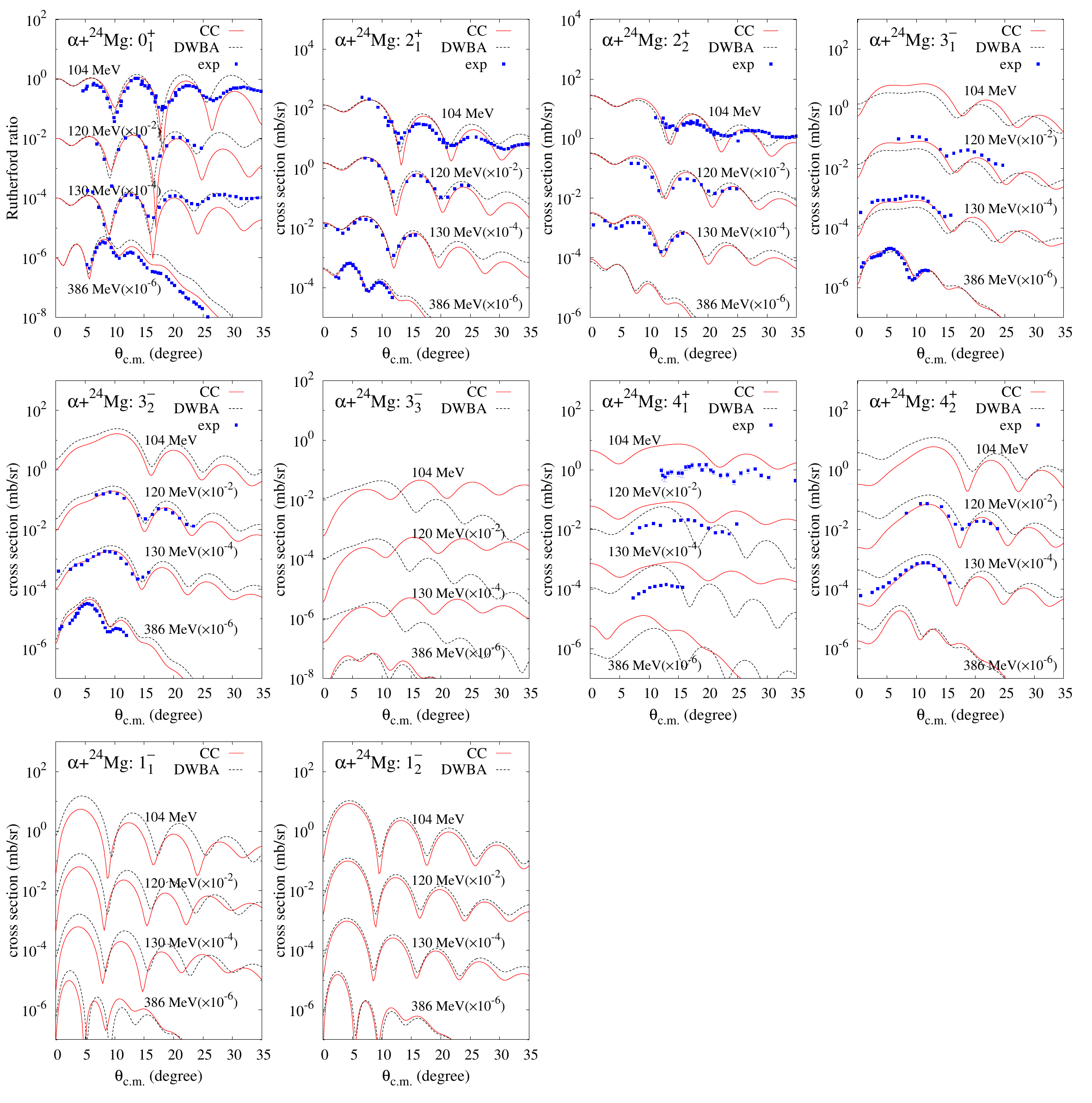}
  \caption{ 
Cross sections of $\alpha$ scattering off $\Mg$ at incident energies of $E_\alpha=104$, 120~MeV, 130~MeV, and 
386 MeV, as calculated with 
MCC+AMD (solid lines labeled ``CC'') and DWBA (dotted lines labeled ``DWBA'').
The experimental data are cross sections 
at $E_\alpha=$104~MeV \cite{Rebel:1972nip,Otuka:2014wzu}, 120~MeV\cite{VanDerBorg:1981qiu}, 130 MeV~\cite{Adachi:2018pql}, and 
386 MeV \cite{Adachi:2018pql}. 
  \label{fig:cross-mg24a}}
\end{figure*}
%%%%%%%%%%%%%%%%%%%%%%%%%

\section{Proton and $\alpha$ scattering: MCC+AMD results}  \label{sec:results2}

The MCC+AMD calculations are performed for proton and $\alpha$ 
scattering using the calculated diagonal and renormalized transition densities.
Our major interest is in extracting structural information, 
particularly to confirm the  band assignment of the negative-parity states 
via reaction analysis of the inelastic cross sections. 
We intend to determine 
how the state dependence of the transition densities affects the cross sections, and 
whether  inelastic scattering can probe the properties of 
three kinds of negative-parity excitations. 

We calculate the elastic and inelastic cross sections of 
proton scattering at incident energies of $E_p=40$~MeV, $49$~MeV, 65~MeV, and $100$~MeV, as well as
$\alpha$ scattering at $E_\alpha=104$~MeV, 120~MeV, 130~MeV, and 386~MeV, and compare the results
with the existing data. 
To see the  CC effects, the one-step calculation of 
the distorted wave born approximation (DWBA) is also performed using the same 
inputs.

\subsection{Proton scattering off $\Mg$}
The calculated cross sections of proton scattering are shown  
in Fig.~\ref{fig:cross-mg24p} and compared with the experimental data. 
The MCC+AMD calculation reasonably reproduces amplitudes of 
the proton elastic and inelastic cross sections in this energy region except for the $4^+_1$ cross sections. 
It also qualitatively describes the diffraction patterns of the cross sections,
though it is not precise enough to reproduce the dip structures of low-energy-backward
and high-energy scattering, 
mainly because the spin-orbit potentials are ignored in the reaction calculation.
Quantitatively, the calculation somewhat underestimates the amplitudes of the $3^-_1$ and $4^+_2$ cross sections. 

Let us discuss the state dependence of the proton-scattering cross sections for the $3^-_1(K^\pi=3^-)$ and 
$3^-_2(K^\pi=0^-)$  states. Both $3^-$ states have strong $\lambda=3$ transitions from the 
ground state, and are therefore strongly populated through inelastic scattering. 
However, as observed in the $(e,e')$ experiment, the two $3^-$ states represent the shape difference
of the form factors because they have different origins of the excitation modes.
%Also in the proton scattering,
%a similar difference in the angular distribution of the $3^-$ cross sections has been 
%reported in the experiment at $E_p=100$~MeV.
Comparing the calculated cross sections of the $3^-_1$ and $3^-_2$ states, 
one can see a difference in the ratio of the first- and second-peak amplitudes;
the second peak of the $3^-_2$ cross sections is suppressed at $E_p=386$~MeV.
This suppression at the second peak for the $3^-_2$ state can be understood 
by the exotic character of the $0^+_1\to 3^-_2$ transition density having 
a nodal structure with the enhanced outer amplitudes shown in Fig.~\ref{fig:trans}(d).
In the $(p,p')$ data at $E_p=100$~MeV, the $3^-_2$ cross sections
fall rapidly compared with the $3^-_1$ state and support the second-peak suppression 
of the calculated cross sections for the $3^-_2(K^\pi=0^-$) state. 
A similar trend is also seen in the $(p,p')$ data at $E_p=65$~MeV, but  
the difference between the $3^-_1$ and $3^-_2$ cross sections is not clearly seen at $E_p=40$~MeV,
for which the correspondence between transition densities and cross sections is not as direct as high
energies because of distortion effects.

In the comparison of the CC results with the one-step~(DWBA) cross sections of proton scattering,
the CC effects for the first and second peaks are found to be minor, except at the $1^-_1$ and $4^+_1$ cross sections. 
The $1^-_1$($K^\pi=0^-$) cross sections are strongly affected by 
the CC effect mainly because of the strong in-band $\lambda=2$ transition 
of the $1^-_1$-$3^-_2$ coupling in the $K^\pi=0^-$ band.
This CC effect suppresses the forward cross sections in the $\theta \le 20^\circ$ region. 
Hence, the calculation describes the enhanced cross sections in the $\theta=20$--$40^\circ$ region 
observed in $E_p=40$~MeV and 65~MeV proton scattering. On the other hand, 
for the $1^-_2$($K^\pi=1^-$) cross sections,  
such CC effects are not significant, even though the 
in-band $\lambda=2$ transition in the $K^\pi=1^-$ band is as large as that in the $K^\pi=0^-$ band.
The weak CC effect for the $1^-_2$($K^\pi=1^-$) state results from the fact 
that the inelastic transition $0^+_1\to 3^-_3$($K^\pi=1^-$) is 
weak compared with the $0^+_1\to 3^-_2$($K^\pi=0^-$) transition.

Thus, we can argue that the present assignments of the $3^-_1$(7.62~MeV) to the $K^\pi=3^-$ band
and the   $1^-_1$(7.56~MeV) and $3^-_2$(8.36~MeV) states to the $K^\pi=0^-$ band are 
supported by the observed proton-scattering cross sections.

%For the $4^+_1$ cross sections of proton scattering, 
%the CC result is unsatisfactory for both the absolute 
%amplitudes and the diffraction patterns because the present AMD calculation fails to
%reproduce the charge form factors, i.e., the transition density of the $4^+_1$ state. 
%Improvement of the structure calculation is needed to solve this problem.}

\subsection{$\alpha$ scattering off $\Mg$}
The results for $\alpha$ scattering are shown in Fig.~\ref{fig:cross-mg24a}.
The MCC+AMD calculation successfully reproduces the observed elastic and inelastic cross sections of
 $\alpha$ scattering
in the energy range of $E_\alpha=100$--400~MeV except in the $4^+_1$ state.
% and the second peak of the $3^-_1$ cross sections at $E_\alpha=120$~MeV.
In the comparison with the one-step cross sections, the CC effects in the $\alpha$-scattering cross sections 
are generally significant, except for the forward cross sections of the $0^+_1$, $2^+_1$, $2^+_2$, and $1^-_2$ states. 
The $4^+_2$ cross sections are hindered by the CC effect of the $\lambda=2$ coupling with the $2^+_1$ and $2^+_2$ states. 
Moreover, the $1^-_1$ cross sections are mainly suppressed because of the $\lambda=2$ coupling with the $3^-_2$ state.
For the $3^-_2$ state, the CC effect somewhat suppresses the cross sections but the effect is 
not as large as in the $1^-_1$ and $4^+_2$ cases. 
On the other hand, the CC effect is opposite for the  $3^-_1$ state;
it enhances the  $3^-_1$ cross sections mainly due to 
the $\lambda=3$ coupling with the $2^+_1$ state and the $\lambda=2$ self-coupling.
As the incident energy increases,   
the CC effects become small but are still non-negligible, even at $E_\alpha=386$~MeV.
The state and energy dependencies of the CC effects found in the calculation 
are essential for reproducing the amplitude of the observed $\alpha$-scattering 
cross sections of 
the $4^+_2$, $3^-_1$, and $3^-_2$ states.

Let us discuss the state dependence of the cross sections of the two $3^-$ states.
In the calculated cross sections for these states, 
a difference is observed in the first-peak shape. 
The $3^-_1$ cross sections show a broad peak, whereas the $3^-_2$ cross sections 
present a narrow peak slightly shifted to backward angles. 
This difference is observed in the experimental data over a wide energy range from 
$E_\alpha=120$ to $386$~MeV. 
This result for the $\alpha$ scattering supports again the assignment of the band-head states of the 
$K^\pi=3^-$ and $K^\pi=0^-$ bands to the $3^-_1$(7.62~MeV) and  $3^-_2$(8.36~MeV) states.

For $E_\alpha=$120 and 130~MeV $\alpha$ scattering, 
the reproduction of the $3^-_1$ cross sections around the second dip ($\theta\sim 15^\circ$) 
is not satisfactory.
This may be explained by 
higher-order effects that are not considered in the present structure and reaction calculations; 
these should be investigated
going forward.
For example, the in-band $\lambda=2$ transition of the $3^-$-$5^-$ coupling is omitted in the
present calculation. Possible mixing between the $K^\pi=3^-_1$ and $K^\pi=0^-$ bands is also expected
from the small energy difference in the observed spectra. 

For the $1^-_1(K^\pi=0^-)$ and $1^-_2(K^\pi=1^-)$ states, 
 no experimental data are available concerning the angular distributions of $\alpha$-scattering 
cross sections in this energy range.  In the calculated cross sections,  
a significant state dependence in the CC effects is found 
between the $1^-_1(K^\pi=0^-)$ and $1^-_2(K^\pi=1^-)$ states.
The strong CC effects are obtained for the $1^-_1(K^\pi=0^-)$ cross sections,
but not for the $1^-_2(K^\pi=1^-)$ cross sections;
this trend for $\alpha$ scattering is similar to that for $p$ scattering. 
For the former state $1^-_1(K^\pi=0^-)$, the strong CC effect arises from the 
two-step processes via the $3^-_2$ state, which include 
the strong $0^+_1\to 3^-_2$~($\lambda=3$) transition and $3^-_2\to 1^-_1$~($\lambda=2$) transitions. 
On the other hand, for the latter state $1^-_2(K^\pi=1^-)$, 
even though the in-band $\lambda=2$ transition between the $1^-_2$ and $3^-_3$ states 
is rather strong, the $0^+_1\to 3^-_3$($\lambda=3$) transition
is weak.
This difference between the $0^+_1\to 3^-_2$ and $0^+_1\to 3^-_3$ transitions results in  
different CC effects in the $1^-_1$ and $1^-_2$ cross sections, which  
can be used as an experimental probe for identifying the two dipole modes, provided that 
the $1^-$ cross sections are measured. 
In particular, possible evidence for the $1^-_1$ state of the $K^\pi=0^-$ band
is the hindered peak amplitude and the dip positions at 
forward angles of the cross sections, as compared with the $1^-_2$ cross sections.

\section{Summary} \label{sec:summary}

The structure and transition properties of the low-lying negative-parity bands 
of $^{24}$Mg were investigated through microscopic structure and reaction calculations 
via proton and $\alpha$ scattering off $^{24}$Mg.

%Structures of $^{20}$Ne were calculated 
%with the VAP version of AMD. 
In the structure calculation for $^{24}$Mg with AMD, 
the $K^\pi=0^+$ ground- and $K^\pi=2^+$ side-bands were obtained by 
triaxial deformation with  
$^{12}\textrm{C}+3\alpha$-like (or $^{12}\textrm{C}+^{12}\textrm{C}$-like) 
structures.
The calculated negative-parity, 
states were classified into the $K^\pi=3^-$,  $K^\pi=0^-$, and $K^\pi=1^-$ bands, 
which are understood as negative-parity excitations 
in the deformed system generated by three types of cluster modes. 
The AMD calculation qualitatively reproduced the observed $E\lambda$-transition strengths and 
elastic and inelastic charge-form factors.

In the MCC+AMD calculation, the AMD transition densities of $\Mg$ were renormalized to 
fit the experimental transition strengths and charge-form factors.
Using the renormalized AMD densities, 
the MCC calculations with 
the Melbourne $g$-matrix $NN$ interaction
were performed for proton and $\alpha$ elastic and inelastic scattering of
energies of $E_p=40$--100 MeV and $E_\alpha=104$--386 MeV. 
The MCC+AMD calculations reasonably reproduced the experimental data for
proton and $\alpha$ elastic and inelastic cross sections in these energy ranges, 
except in the case of the $4^+_1$ cross sections. 

In the reaction analysis of the proton and $\alpha$ inelastic scattering processes using the MCC+AMD calculation, 
the transition properties of the negative-parity bands were discussed. 
Comparison of the calculated  $(p,p')$ and $(\alpha,\alpha')$ cross sections 
as well as the charge form factors with the experimental data
showed that the band-head states of the $K^\pi=3^-$, $K^\pi=0^-$, and $K^\pi=1^-$ bands were assigned to the 
experimental $3^-_1$(7.62~MeV), $1^-_1$(7.56~MeV), and $1^-_2$(8.44~MeV) states, respectively. 
Moreover, 
the $3^-$ member of the $K^\pi=0^-$ band was assigned to the $3^-_2$(8.36~MeV) state. 

The $3^-_1$ and $3^-_2$ states were strongly excited by $p$ and $\alpha$ scattering, and 
one-step processes dominantly contribute to the angular distributions of these cross sections;
however, CC effects remain essential for reproducing the absolute amplitudes of the $(\alpha,\alpha')$ 
cross sections. The shape difference in the $0^-_1\to 3^-$ transition densities between
the $3^-_1$ and $3^-_2$ states can be observed in the first-peak shape 
of the $(p,p')$ and $(\alpha,\alpha')$ cross sections.
For $1^-$ states, the present calculation predicted strong CC effects on the 
$1^-_1$ cross sections and weak CC effects on the $1^-_2$ cross sections.
For the  $1^-_1$ cross sections, the strong in-band $\lambda=2$ transition 
of the $1^-_1$-$3^-_2$ coupling significantly 
changes the first-peak shape of the $(p,p')$ cross sections via the two-step process 
$0^+_1\to 3^-_2\to 1^-_1$, which is supported by proton scattering data. 
It also strongly affects the $1^-_1$ cross sections of $\alpha$ scattering.

The present results prove that 
proton and $\alpha$ inelastic scattering are good probes for investigating
the properties of transitions and the band structure of excited states. 
The MCC approach combining the microscopic structure and reaction calculations was found to be 
a powerful tool for reaction analysis. 

The present AMD calculation 
is the first microscopic structure calculation to 
reproduce the energy ordering of the $K^\pi=0^-$, $K^\pi=1^-$, and $K^\pi=3^-$ bands of $^{24}$Mg.
However, it has problems in precisely reproducing
the structural properties. For example, the calculation generally underestimated their transition strengths
and overestimated their excitation energies,  
although it reproduced the energy ordering of the excited bands.
These issues remain to be solved by improvement of the structure calculation.
The present calculation also failed to reproduce the shape of the $0^+_1\to 4^+_1$ form factors, 
which was a crucial problem for describing the observed $(p,p')$ and $(\alpha,\alpha')$ cross sections for the
$4^+_1$ state. 
We can solve this problem with another version of the AMD calculation that yields 
superior results for the excitation energies and transition properties of the 
$K^\pi=0^+$ and $K^\pi=2^+$ bands. In a future paper, 
we will present a detailed investigation of proton and $\alpha$ inelastic scattering 
to the $4^+_1$ state using the MCC+AMD approach
with improved AMD densities for $\Mg$. 

\begin{acknowledgments}
The computational calculations of this work were performed using the
supercomputer at the Yukawa Institute for Theoretical Physics at Kyoto University. The work was supported
by Grants-in-Aid of the Japan Society for the Promotion of Science (Grant Nos. JP18K03617, JP16K05352, and 18H05407) and by a grant of the joint research project of the Research Center for Nuclear Physics at Osaka
University.
\end{acknowledgments}


\begin{thebibliography}{9}

\bibitem{Branford:1975ciy}
D.~Branford, A.~C.~McGough and I.~F.~Wright,
%``Lifetime and branching measurements on the K π = 0 + and K π = 2 + 24 Mg rotational levels,''
Nucl. Phys. A \textbf{241}, 349-364 (1975).
%doi:10.1016/0375-9474(75)90324-3
%37 citations counted in INSPIRE as of 02 Sep 2020

\bibitem{Fifield:1979gfv}
L.~K.~Fifield, E.~F.~Garman, M.~J.~Hurst, T.~J.~M.~Symons, F.~Watt, C.~H.~Zimmerman and K.~W.~Allen,
%``Radiative decays of unbound high-spin states in 24 Mg (II),''
Nucl. Phys. A \textbf{322}, 1-12 (1979).
%doi:10.1016/0375-9474(79)90329-4
%19 citations counted in INSPIRE as of 01 Sep 2020

\bibitem{Keinonen:1989ltz}
J.~Keinonen, P.~Tikkanen, A.~Kuronen, Á.~Z.~Kiss, E.~Somorjai and B.~H.~Wildenthal,
%``Short lifetimes in 24 Mg for test of rotational collectivity in shell-model wave functions,''
Nucl. Phys. A \textbf{493}, 124-144 (1989).
%doi:10.1016/0375-9474(89)90536-8
%16 citations counted in INSPIRE as of 01 Sep 2020

\bibitem{Horikawa:1971oau}
Y.~Horikawa, Y.~Torizuka, A.~Nakada, S.~Mitsunobu, Y.~Kojima and M.~Kimura,
%``The deformations in 20 Ne, 24 Mg and 28 Si from electron scattering,''
Phys. Lett. B \textbf{36}, 9-11 (1971).
%doi:10.1016/0370-2693(71)90306-6
%77 citations counted in INSPIRE as of 02 Sep 2020

\bibitem{Horikawa1972}
Y.~Horikawa,
%``Deformations of Ground-Bands in Ne20, Mg24 and Si28'' I
Prog.\ Theor.\ Phys.\ {\bf 47}, 867 (1072).  

\bibitem{Nakada:1972}
A.~Nakada and Y.~ Torizuka, J. Phys. Soc. Jpn {\bf 32} , 1 (1972). 

\bibitem{Johnston_1974}
{A Johnston and T E Drake}, J.\ Phys. A7, 898 (1974).
%	doi = {10.1088/0305-4470/7/8/004},
%	url = {https://doi.org/10.1088%2F0305-4470%2F7%2F8%2F004},
%{A study of24Mg by inelastic electron scattering},

\bibitem{Li:1974vj}
G.~C.~Li, I.~Sick and M.~R.~Yearian,
%``High Momentum Transfer electron Scattering from Mg-24, Al-27, Si-28 and S-32,''
Phys. Rev. C \textbf{9}, 1861 (1974).
%doi:10.1103/PhysRevC.9.1861
%102 citations counted in INSPIRE as of 02 Sep 2020

\bibitem{Zarek:1978cvz}
H.~Zarek, S.~Yen, B.~O.~Pich, T.~E.~Drake, C.~F.~Williamson, S.~Kowalski, C.~P.~Sargent, W.~Chung, B.~H.~Wildenthal, M.~Harvey and H.~C.~Lee,
%``Electroexcitation and the determination of the K -band structure in 24 Mg,''
Phys. Lett. B \textbf{80}, 26-29 (1978).
%doi:10.1016/0370-2693(78)90297-6
%18 citations counted in INSPIRE as of 01 Sep 2020

\bibitem{Zarek:1984fm}
H.~Zarek, S.~Yen, B.~O.~Pich, T.~E.~Drake, C.~F.~Williamson, S.~Kowalski and C.~P.~Sargent,
%``INELASTIC ELECTRON SCATTERING TO NEGATIVE PARITY STATES OF MG-24,''
Phys. Rev. C \textbf{29}, 1664-1671 (1984).
%doi:10.1103/PhysRevC.29.1664
%5 citations counted in INSPIRE as of 01 Sep 2020



\bibitem{Blanpied:1990vd}
G.~S.~Blanpied, J.~Hernandez, C.~S.~Mishra, W.~K.~Mize, C.~S.~Whisnant, B.~G.~Ritchie, C.~L.~Morris, S.~J.~Seestrom- Morris, C.~F.~Moore, P.~A.~Seidl, R.~A.~Lindgren, B.~H.~Wildenthal and R.~A.~Gilman,
%``Pion elastic and inelastic scattering from Mg-24 and Mg-26,''
Phys. Rev. C \textbf{41}, 1625-1636 (1990).
%doi:10.1103/PhysRevC.41.1625
%5 citations counted in INSPIRE as of 01 Sep 2020

\bibitem{Haouat:1984zz}
G.~Haouat, C.~Lagrange, R.~de Swiniarski, F.~Dietrich, J.~P.~Delaroche and Y.~Patin,
%``Nuclear deformations of Mg-24, Si-28, and S-32 from fast neutron scattering,''
Phys. Rev. C \textbf{30}, 1795-1809 (1984).
%doi:10.1103/PhysRevC.30.1795
%17 citations counted in INSPIRE as of 02 Sep 2020

\bibitem{Rush:1967zwr}
A.~A.~Rush, E.~J.~Burge, V.~E.~Lewis, D.~A.~Smith and N.~K.~Ganguly,
%``Elastic and inelastic scattering of 49.5 MeV protons by 24 Mg,''
Nucl. Phys. A \textbf{104}, 340-352 (1967).
%doi:10.1016/0375-9474(67)90561-1
%27 citations counted in INSPIRE as of 01 Sep 2020

\bibitem{Rush:1968qwc}
A.~A.~Rush and N.~K.~Ganguly,
%``Collective-model analysis of the inelastic scattering of 49.5 MeV protons by 24 Mg,''
Nucl. Phys. A \textbf{117}, 101-112 (1968).
%doi:10.1016/0375-9474(68)90863-4
%22 citations counted in INSPIRE as of 01 Sep 2020

\bibitem{Zwieglinski:1978zza}
B.~Zwieglinski, G.~M.~Crawley, H.~Nann and J.~A.~Nolen,
%``Inelastic scattering of 40 MeV protons from Mg-24. 1. Natural parity transitions,''
Phys. Rev. C \textbf{17}, 872-887 (1978).
%doi:10.1103/PhysRevC.17.872
%35 citations counted in INSPIRE as of 01 Sep 2020

\bibitem{Zwieglinski:1978zz}
B.~Zwieglinski, G.~M.~Crawley, W.~Chung, H.~Nann and J.~A.~Nolen,
%``Inelastic scattering of 40 MeV protons from Mg-24. 2. Microscopic calculations for positive parity states,''
Phys. Rev. C \textbf{18}, 1228-1236 (1978).
%doi:10.1103/PhysRevC.18.1228
%12 citations counted in INSPIRE as of 01 Sep 2020

\bibitem{Kato:1985zz}
S.~Kato, K.~Okada, M.~Kondo, K.~Hosono, T.~Saito, N.~Matsuoka, K.~Hatanaka, T.~Noro, S.~Nagamachi, H.~Shimizu, K.~Ogino, Y.~Kadota, S.~Matsuki and M.~Wakai,
%``Inelastic scattering of 65 MeV protons from C-12, Mg-24, Si-28, and S-32,''
Phys. Rev. C \textbf{31}, 1616-1632 (1985).
%doi:10.1103/PhysRevC.31.1616
%42 citations counted in INSPIRE as of 24 Sep 2020


\bibitem{Horowitz:1969eso}
Y.~S.~Horowitz, N.~K.~Sherman and R.~E.~Bell,
%``Inelastic scattering of 100 MeV protons from Mg and Si using a Ge(Li) total absorption proton counter,''
Nucl. Phys. A \textbf{134}, 577-598 (1969).
%doi:10.1016/0375-9474(69)90023-2
%12 citations counted in INSPIRE as of 01 Sep 2020


\bibitem{Lombard:1978zz}
R.~M.~Lombard, J.~L.~Escudie and M.~Soyeur,
%``Distorted-wave Born approximation and coupled channel analyses of low-energy polarized proton scattering on Mg-24,''
Phys. Rev. C \textbf{18}, 42-55 (1978).
%doi:10.1103/PhysRevC.18.42
%19 citations counted in INSPIRE as of 02 Sep 2020



\bibitem{Ray:1979zza}
L.~Ray, G.~S.~Blanpied and W.~R.~Coker,
%``Coupled-channels analysis of proton inelastic scattering to the gamma-vibrational band in Mg-24,''
Phys. Rev. C \textbf{20}, 1236-1243 (1979).
%doi:10.1103/PhysRevC.20.1236
%18 citations counted in INSPIRE as of 01 Sep 2020


\bibitem{Blanpied:1979im}
G.~Blanpied, N.~M.~Hintz, G.~S.~Kyle, J.~W.~Palm, R.~Liljestrand, M.~Barlett, C.~Harvey, G.~W.~Hoffmann, L.~Ray and D.~G.~Madland,
%``PROTON SCATTERING FROM MG-24 AT 0.8-GEV,''
Phys. Rev. C \textbf{20}, 1490-1497 (1979).
%doi:10.1103/PhysRevC.20.1490
%18 citations counted in INSPIRE as of 01 Sep 2020


\bibitem{DeLeo:1981zz}
R.~De Leo, G.~D'Erasmo, A.~Pantaleo, M.~N.~Harakeh, S.~Micheletti and M.~Pignanelli,
%``Proton excitation of Mg-24, Mg-26, and gamma bands,''
Phys. Rev. C \textbf{23}, 1355-1363 (1981).
%doi:10.1103/PhysRevC.23.1355
%15 citations counted in INSPIRE as of 21 Aug 2020


\bibitem{Amos:1984aph}
K.~Amos and W.~Bauhoff,
%``Analyses of electron and proton scattering to low-excitation isoscalar states in 20 Ne, 24 Mg and 28 Si,''
Nucl. Phys. A \textbf{424}, 60-80 (1984).
%doi:10.1016/0375-9474(84)90128-3
%9 citations counted in INSPIRE as of 02 Sep 2020


\bibitem{Griffiths:1967hrr}
R.~J.~Griffiths,
%``Elastic and inelastic scattering of 29 MeV 3 He by 24 Mg,''
Nucl. Phys. A \textbf{102}, 329-336 (1967).
%doi:10.1016/0375-9474(67)90026-7
%9 citations counted in INSPIRE as of 02 Sep 2020


\bibitem{VanDerBorg:1979pzv}
K.~Van Der Borg, M.~N.~Harakeh and B.~S.~Nilsson,
%``Excitation of ground and gamma bands in the 24 Mg(α, α') 24 Mg reaction at 120 MeV,''
Nucl. Phys. A \textbf{325}, 31-44 (1979).
%doi:10.1016/0375-9474(79)90149-0
%24 citations counted in INSPIRE as of 01 Sep 2020


\bibitem{Naqib1968}
I. M. Naqib and J. S. Blair, 
%``Inelastic Scattering of 42-MeV Alpha Particles by Mg24''
Phys. Rev. {\bf 165}, 1250 (1968).


\bibitem{Rebel:1972nip} 
  H.~Rebel, G.~W.~Schweimer, G.~Schatz, J.~Specht, R.~Löhken, G.~Hauser, D.~Habs and H.~Klewe-Nebenius,
  %``Quadrupole and hexadecapole deformation of 2s-1d shell nuclei,''
  Nucl.\ Phys.\ A {\bf 182}, 145 (1972).
%  doi:10.1016/0375-9474(72)90207-2
  %%CITATION = doi:10.1016/0375-9474(72)90207-2;%%
  %110 citations counted in INSPIRE as of 17 Feb 2020


\bibitem{VanDerBorg:1981qiu} 
  K.~Van Der Borg, M.~N.~Harakeh and A.~Van Der Woude,
  %``The isoscalar strength distribution in 24, 26 Mg, 28 Si and 40 Ca obtained from inelastic alpha scattering at 120 Mev,''
  Nucl.\ Phys.\ A {\bf 365}, 243 (1981).
%  doi:10.1016/0375-9474(81)90297-9
  %%CITATION = doi:10.1016/0375-9474(81)90297-9;%%
  %69 citations counted in INSPIRE as of 19 Dec 2019


\bibitem{Adachi:2018pql}
S.~Adachi, T.~Kawabata, K.~Minomo, T.~Kadoya, N.~Yokota, H.~Akimune, T.~Baba, H.~Fujimura, M.~Fujiwara, Y.~Funaki, T.~Furuno, T.~Hashimoto, K.~Hatanaka, K.~Inaba, Y.~Ishii, M.~Itoh, C.~Iwamoto, K.~Kawase, Y.~Maeda, H.~Matsubara, Y.~Matsuda, H.~Matsuno, T.~Morimoto, H.~Morita, M.~Murata, T.~Nanamura, I.~Ou, S.~Sakaguchi, Y.~Sasamoto, R.~Sawada, Y.~Shimizu, K.~Suda, A.~Tamii, Y.~Tameshige, M.~Tsumura, M.~Uchida, T.~Uesaka, H.~P.~Yoshida and S.~Yoshida,
%``Systematic analysis of inelastic α scattering off self-conjugate A=4n nuclei,''
 %``Systematic analysis of inelastic α scattering off self-conjugate A=4n nuclei,''
  Phys.\ Rev.\ C {\bf 97},  014601 (2018).
%  doi:10.1103/PhysRevC.97.014601
  %%CITATION = doi:10.1103/PhysRevC.97.014601;%%
  %1 citations counted in INSPIRE as of 07 Jan 2019



\bibitem{Nesterenko:2017rcc}
V.~O.~Nesterenko, A.~Repko, J.~Kvasil and P.~G.~Reinhard,
%``Individual Low-Energy Toroidal Dipole State in $^{24}$Mg,''
Phys. Rev. Lett. \textbf{120}, no.18, 182501 (2018).
%doi:10.1103/PhysRevLett.120.182501
%[arXiv:1711.08953 [nucl-th]].
%12 citations counted in INSPIRE as of 02 Sep 2020

\bibitem{Nesterenko:2019dnt}
V.~O.~Nesterenko, A.~Repko, J.~Kvasil and P.~G.~Reinhard,
%``Individual dipole toroidal states: Main features and search in the $(e,e′)$ reaction,''
Phys. Rev. C \textbf{100}, no.6, 064302 (2019).
%doi:10.1103/PhysRevC.100.064302
%[arXiv:1904.08302 [nucl-th]].
%2 citations counted in INSPIRE as of 02 Sep 2020

\bibitem{Kimura2012}
M. Kimura, R. Yoshida, M. Isaka, 
Prog.\ Theor.\ Phys.\ {\bf 127},287 (2012).
%DOI: 10.1143/PTP.127.287

\bibitem{Chiba:2019dap}
Y.~Chiba, Y.~Kanada-En'yo and Y.~Shikata,
%``Cluster correlation and nuclear vorticity in low-lying $1^-$ states of $^{24}$Mg,''
[arXiv:1911.08734 [nucl-th]].
%0 citations counted in INSPIRE as of 02 Sep 2020

\bibitem{KanadaEnyo:1994kw} 
  Y.~Kanada-Enyo and H.~Horiuchi,
  %``Clustering in yrast states of Ne-20 studied with antisymmetrized molecular dynamics,''
  Prog.\ Theor.\ Phys.\  {\bf 93}, 115 (1995).
%  doi:10.1143/PTP.93.115
  %%CITATION = doi:10.1143/PTP.93.115;%%
  %81 citations counted in INSPIRE as of 17 Feb 2020

\bibitem{KanadaEnyo:1995tb}
  Y.~Kanada-En'yo, H.~Horiuchi and A.~Ono,
  %``Structure of Li and Be isotopes studied with antisymmetrized molecular
  %dynamics,''
  Phys.\ Rev.\  C {\bf 52}, 628  (1995).
  %%CITATION = PHRVA,C52,628;%%


\bibitem{KanadaEn'yo:1998rf}
  Y.~Kanada-En'yo,
  %``Variation after angular momentum projection for the study of excited states based on antisymmetrized molecular dynamics,''
  Phys.\ Rev.\ Lett.\  {\bf 81}, 5291 (1998).
%  doi:10.1103/PhysRevLett.81.5291
%  [nucl-th/0204039].
  %%CITATION = doi:10.1103/PhysRevLett.81.5291;%%
  %86 citations counted in INSPIRE as of 06 Dec 2015

\bibitem{KanadaEn'yo:2012bj}
  Y.~Kanada-En'yo, M.~Kimura and A.~Ono,
  %``Antisymmetrized molecular dynamics and its applications to cluster phenomena,''
  Prog. Theor. Exp. Phys.  {\bf 2012}  01A202 (2012).
%  [arXiv:1202.1864 [nucl-th]].
  %%CITATION = ARXIV:1202.1864;%%
  %1 citations counted in INSPIRE as of 28 Jul 2013

\bibitem{Kanada-Enyo:2019prr}
  Y.~Kanada-En'yo and K.~Ogata,
  %``$\alpha$ scattering cross sections on $^{12}$C with microscopic coupled-channel calculation,''
  Phys.\ Rev.\ C {\bf 99}, no. 6, 064601 (2019).
%  doi:10.1103/PhysRevC.99.064601
%  [arXiv:1903.10164 [nucl-th]].
  %%CITATION = doi:10.1103/PhysRevC.99.064601;%%

\bibitem{Kanada-Enyo:2019qbp}
  Y.~Kanada-En'yo and K.~Ogata,
  %``First microscopic coupled-channels calculation of cross sections for inelastic $\alpha$ scattering off $^{16}O,''
  Phys.\ Rev.\ C {\bf 99}, no. 6, 064608 (2019).
%  doi:10.1103/PhysRevC.99.064608
%  [arXiv:1904.03811 [nucl-th]].
  %%CITATION = doi:10.1103/PhysRevC.99.064608;%%


\bibitem{Kanada-Enyo:2019uvg} 
  Y.~Kanada-En'yo and K.~Ogata,
  %``Microscopic calculation of inelastic proton scattering off $^{18}$O, $^{10}$Be, $^{12}$Be, and $^{16}$C for study of neutron excitation in neutron-rich nuclei,''
  Phys.\ Rev.\ C {\bf 100}, no. 6, 064616 (2019).
%  doi:10.1103/PhysRevC.100.064616
%  [arXiv:1908.03293 [nucl-th]].
  %%CITATION = doi:10.1103/PhysRevC.100.064616;%%

\bibitem{Kanada-Enyo:2020zpl}
Y.~Kanada-En'yo and K.~Ogata,
%``Transition properties of low-lying states in $^{28}$Si probed via inelastic proton and $\alpha$ scattering,''
Phys. Rev. C \textbf{101}, no.6, 064607 (2020).
%doi:10.1103/PhysRevC.101.064607
%[arXiv:2002.02625 [nucl-th]].
%3 citations counted in INSPIRE as of 14 Sep 2020

\bibitem{Kanada-Enyo:2020goh}
Y.~Kanada-En'yo and K.~Ogata,
%``Properties of $K^\pi=0^+_1$, $K^\pi=2^-$, and $K^\pi=0^-_1$ bands of $^{20}$Ne probed via proton and alpha inelastic scattering,''
Phys. Rev. C \textbf{101}, no.6, 064308 (2020).
%doi:10.1103/PhysRevC.101.064308
%[arXiv:2003.14076 [nucl-th]].
%0 citations counted in INSPIRE as of 14 Sep 2020

\bibitem{Ogata:2020umn}
K.~Ogata, Y.~Chiba and Y.~Sakuragi,
%``Correspondence between $\alpha$ inelastic cross sections off $^{24}$Mg and isoscalar monopole strengths,''
[arXiv:2001.09627 [nucl-th]].
%0 citations counted in INSPIRE as of 11 Sep 2020


\bibitem{Amos:2000}
K. Amos, P. J. Dortmans, H. V. von Geramb, S. Karataglidis, and J. Raynal,
Adv.~Nucl.~Phys. {\bf 25}, 275 (2000).

\bibitem{Mac87}
R. Machleidt, K. Holinde, and Ch. Elster, Phys. Reports {\bf 149}, 1 (1987).




\bibitem{Kanada-Enyo:1999bsw}
  Y.~Kanada-En'yo, H.~Horiuchi and A.~Dote,
  %``Structure of excited states of Be-10 studied with antisymmetrized molecular dynamics,''
  Phys.\ Rev.\ C {\bf 60}, 064304 (1999).
%  doi:10.1103/PhysRevC.60.064304
% [nucl-th/9905048].
  %%CITATION = doi:10.1103/PhysRevC.60.064304;%%
  %112 citations counted in INSPIRE as of 02 Jun 2019

\bibitem{Kanada-Enyo:2003fhn}
  Y.~Kanada-En'yo and H.~Horiuchi,
  %``Cluster structures of the ground and excited states of Be-12 studied with antisymmetrized molecular dynamics,''
  Phys.\ Rev.\ C {\bf 68}, 014319 (2003).
%  doi:10.1103/PhysRevC.68.014319
%  [nucl-th/0301059].
  %%CITATION = doi:10.1103/PhysRevC.68.014319;%%
  %68 citations counted in INSPIRE as of 02 Jun 2019


\bibitem{TOHSAKI}
 T. Ando, K.Ikeda, and A. Tohsaki, Prog. Theor. Phys.
 {\bf 64}, 1608 (1980).
\bibitem{LS1}
 R. Tamagaki, Prog. Theor. Phys. {\bf 39}, 91 (1968).

\bibitem{LS2}
 N. Yamaguchi, T. Kasahara, S. Nagata, and Y. Akaishi,
 Prog. Theor. Phys. {\bf 62}, 1018 (1979).

\bibitem{Egashira:2014zda}
  K.~Egashira, K.~Minomo, M.~Toyokawa, T.~Matsumoto and M.~Yahiro,
  %``Microscopic optical potentials for $^4$He scattering,''
  Phys.\ Rev.\ C {\bf 89}, 064611 (2014).
%  doi:10.1103/PhysRevC.89.064611
%  [arXiv:1404.2735 [nucl-th]].
  %%CITATION = doi:10.1103/PhysRevC.89.064611;%%
  %14 citations counted in INSPIRE as of 14 Feb 2019

\bibitem{Minomo:2009ds}
K.~Minomo, K.~Ogata, M.~Kohno, Y.~R.~Shimizu and M.~Yahiro,
%``The Brieva-Rook Localization of the Microscopic Nucleon-Nucleus Potential,''
J. Phys. G \textbf{37}, 085011 (2010).
%doi:10.1088/0954-3899/37/8/085011
%[arXiv:0911.1184 [nucl-th]].
%40 citations counted in INSPIRE as of 13 Sep 2020

\bibitem{Toyokawa:2013uua}
M.~Toyokawa, K.~Minomo and M.~Yahiro,
%``Mass-number and isotope dependence of local microscopic optical potentials for polarized proton scattering,''
Phys. Rev. C \textbf{88}, no.5, 054602 (2013).
%doi:10.1103/PhysRevC.88.054602
%[arXiv:1304.7884 [nucl-th]].
%29 citations counted in INSPIRE as of 13 Sep 2020


\bibitem{Minomo:2016hgc}
  K.~Minomo and K.~Ogata,
  %``Consistency between the monopole strength of the Hoyle state determined by structural calculation and that extracted from reaction observables,''
  Phys.\ Rev.\ C {\bf 93}, 051601(R) (2016).


\bibitem{Minomo:2017hjl}
  K.~Minomo, K.~Washiyama and K.~Ogata,
  %``Reexamination of microscopic optical potentials based on multiple scattering theory,''
  arXiv:1712.10121 [nucl-th].
  %%CITATION = ARXIV:1712.10121;%%
  %1 citations counted in INSPIRE as of 02 Jun 2019}.



\bibitem{Firestone:2007crk}
R.~B.~Firestone,
%``Nuclear Data Sheets for A = 24,''
Nucl. Data Sheets \textbf{108}, 2319-2392 (2007).
%doi:10.1016/j.nds.2007.10.001
%54 citations counted in INSPIRE as of 01 Sep 2020


\bibitem{Angeli2013}
I.~Angeli and K.~P.~Marinova, At.~Data Nucl.~Data Tables {\bf 99}, 69 (2013).



\bibitem{KIBEDI:2002tqu}
T.~Kibédi and R.~h.~Spear,
%``REDUCED ELECTRIC-OCTUPOLE TRANSITION PROBABILITIES, B ( E 3;0 1 + →3 1 − )—AN UPDATE,''
Atom. Data Nucl. Data Tabl. \textbf{80}, 35-82 (2002).
%doi:10.1006/adnd.2001.0871
%239 citations counted in INSPIRE as of 01 Sep 2020


\bibitem{Otuka:2014wzu}
%\bibitem{EXFOR}
N.~Otuka, E.~Dupont, V.~Semkova, B.~Pritychenko, A.~I.~Blokhin, M.~Aikawa, S.~Babykina, M.~Bossant, G.~Chen, S.~Dunaeva, R.~A.~Forrest, T.~Fukahori, N.~Furutachi, S.~Ganesan, Z.~Ge, O.~O.~Gritzay, M.~Herman, S.~Hlavač, K.~Katō, B.~Lalremruata, Y.~O.~Lee, A.~Makinaga, K.~Matsumoto, M.~Mikhaylyukova, G.~Pikulina, V.~G.~Pronyaev, A.~Saxena, O.~Schwerer, S.~P.~Simakov, N.~Soppera, R.~Suzuki, S.~Takács, X.~Tao, S.~Taova, F.~Tárkányi, V.~V.~Varlamov, J.~Wang, S.~C.~Yang, V.~Zerkin and Y.~Zhuang,
%``Towards a More Complete and Accurate Experimental Nuclear Reaction Data Library (EXFOR): International Collaboration Between Nuclear Reaction Data Centres (NRDC),''
Nucl. Data Sheets \textbf{120}, 272-276 (2014).
%doi:10.1016/j.nds.2014.07.065
%[arXiv:2002.07114 [nucl-ex]].
%170 citations counted in INSPIRE as of 24 Sep 2020




\end{thebibliography}
\end{document}